\documentclass[10pt,letterpaper]{article}

\usepackage{amsmath,amssymb}

\usepackage{changepage}

\usepackage[utf8x]{inputenc}
\usepackage{textcomp}
\usepackage{textgreek}

\usepackage{textcomp,marvosym}

\usepackage{cite}

\usepackage{nameref,hyperref}

\usepackage[right]{lineno}

\usepackage{microtype}
\DisableLigatures[f]{encoding = *, family = * }

\usepackage[table]{xcolor}

\usepackage{array}

\newcolumntype{+}{!{\vrule width 2pt}}

\newlength\savedwidth

\newcommand\thickhline{\noalign{\global\savedwidth\arrayrulewidth\global\arrayrulewidth 2pt}%
\hline
\noalign{\global\arrayrulewidth\savedwidth}}

\usepackage[aboveskip=1pt,labelfont=bf,labelsep=period,justification=raggedright,singlelinecheck=off]{caption}
\renewcommand{\figurename}{Fig}

\makeatletter
\renewcommand{\@biblabel}[1]{\quad#1.}
\makeatother

\usepackage{lastpage,fancyhdr,graphicx}
\usepackage{epstopdf}

\newcommand{\bytes}{\mbox{B}}
\newcommand{\byte}{\mbox{B}}
\newcommand{\second}{\mbox{s}}

\newcommand{\flop}{\mbox{flop}}

\newcommand{\cycle}{\mbox{cy}}
\newcommand{\cycles}{\mbox{cy}}

\newcommand{\CIT}{\mbox{\cycle/\mbox{scalar iter}}}

\newcommand{\BC}{\mbox{\byte/\cycle}}

\newcommand{\GBS}{\mbox{G\byte/\second}}

\newcommand{\GFS}{\mbox{G\flop/\second}}

\newcommand{\us}{\mbox{$\mu$\mbox{s}}}
\newcommand{\usB}{\mbox{$\mu$\mbox{s}/\byte}}

\newcommand{\GB}{\mbox{GB}}
\newcommand{\KB}{\mbox{kB}}
\newcommand{\MB}{\mbox{MB}}

\newcommand{\olsep}{\|}
\newcommand{\nolsep}{|}
\newcommand{\ecmspace}{\,}
\newcommand{\ecm}[6]{\mbox{$\left\{{#1}\ecmspace\olsep\ecmspace {#2}\ecmspace\nolsep\ecmspace {#3}\ecmspace\nolsep\ecmspace {#4}\ecmspace\nolsep\ecmspace {#5}\right\}\ecmspace{#6}$}}
\newcommand{\epsep}{\rceil}
\newcommand{\ecmp}[5]{\mbox{$\left\{{#1}\ecmspace\epsep\ecmspace {#2}\ecmspace\epsep\ecmspace {#3}\ecmspace\epsep\ecmspace {#4}\right\}\ecmspace{#5}$}}

\newcommand{\revision}[2]{#2}

\begin{document}
\vspace*{0.2in}

\begin{flushleft}
{\Large
\textbf\newline{Telling neuronal apples from oranges: analytical performance modeling of neural tissue simulations} 
}
\newline
\\
    Francesco Cremonesi\textsuperscript{1} and
    Felix Sch\"urmann\textsuperscript{1*}
\\
\bigskip
\textbf{1} Blue Brain Project, Brain Mind Institute, \'Ecole polytechnique f\'ed\'erale de Lausanne, Switzerland
\\
\bigskip

%
%





* felix.schuermann@epfl.ch

\end{flushleft}
\section*{Abstract}
Computational modeling and simulation have become essential tools in the quest to better understand the brain’s makeup and to decipher the causal interrelations of its components.
The breadth of biochemical and biophysical processes and structures in the brain has led to the development of a large variety of model abstractions and specialized tools, often times requiring high performance computing resources for their timely execution.
What has been missing so far was an in-depth analysis of the complexity of the computational kernels, hindering a systematic approach to identifying bottlenecks of algorithms and hardware, and their specific combinations.
If whole brain models are to be achieved on emerging computer generations, models and simulation engines will have to be carefully co-designed for the intrinsic hardware tradeoffs.
For the first time, we present a systematic exploration based on analytic performance modeling.
We base our analysis on three in silico models, chosen as representative examples of the most widely employed modeling abstractions: current-based point neurons, conductance-based point neurons and conductance-based detailed neurons.
We identify that the synaptic modeling formalism, i.e.
current or conductance-based representation, and not the level of morphological detail, is the most significant factor in determining the properties of memory bandwidth saturation and shared-memory scaling of in silico models.
Even though general purpose computing has, until now, largely been able to deliver high performance, we find that for all types of abstractions, network latency and memory bandwidth will become severe bottlenecks as the number of neurons to be simulated grows.
\revision{The methods used in this paper}{By adapting and extending a performance modeling approach, we} deliver a first characterization of the performance landscape of brain tissue simulations, allowing us to pinpoint current bottlenecks for state-of-the-art in silico models, and make projections for future hardware and software requirements.
\section*{Introduction and motivation}
In the field of computational neuroscience, simulations of biological neural networks represent one of the fundamental tools for hypothesis testing and exploration.
A widely used scale of representation are neuron-based approaches, i.e. models of brain tissue in which the fundamental unit is represented by a neuronal cell.
This representation is important as it allows for a faithful matching of the model with a range of anatomical and electrophysiological data~\cite{markram2015reconstruction,potjans2012cell,pozzorini2015automated,hagen2016hybrid,hagen2018multimodal}.
While determining the adequate level of detail is a formidable challenge with respect to the system modeled, so is the addressing of an efficient implementation of the simulation in software. The size of such networks, both in terms of number of neurons and \revision{number}{synapses and rate} of synaptic events, as well as the level of biological detail required to answer meaningful questions about the brain, mean that these simulations come at a large computational cost. 

\revision{
A second approach that in silico neuroscientists have used}
{One axis of exploration} to mitigate the computational cost of brain tissue simulation has been to increase the number of compute nodes and benefit from the improved performance of individual processors over time, leveraging a constant trend of ever more powerful computing capabilities of microprocessors dictated by Moore's law.
Methods that exploit the distributed nature of computer clusters \revision{
, as well as the shared memory parallelism capabilities of modern multicore processors, were introduced in~\cite{migliore2006parallel,eichner2009neural,plesser2007efficient} for some of the most widely used simulation software packages
}{
  were first introduced in~\cite{morrison2005advancing} for point neurons and~\cite{migliore2006parallel} for morphologically detailed models. Such techniques were then refined and extended to exploit shared-memory parallelism in~\cite{eichner2009neural,plesser2007efficient}
}, while recently developed simulators such as e.g. CoreNEURON~\cite{kumbhar2019coreneuron}, Auryn~\cite{zenke2014limits}, Arbor~\cite{klijn2017arbor} have these fundamental capabilities built-in from the beginning.

\revision{One approach is}{A second approach has been} to design efficient \revision{}{numerical} algorithms and software able to optimize resource usage.
A review of numerical methods with a focus on computational neuroscience was given in~\cite{mascagni1989numerical}, \revision{
As an example, the seminal work~\cite{hines1984efficient} introduced a time-splitting numerical}{
while a staggered time integration} method and a neuron representation scheme that allow to solve a large class of morphologically detailed neuron models in linear complexity \revision{}{were introduced in~\cite{hines1984efficient}}.
\revision{}{
Time integration techniques that allow adaptation of the timestep to maximise accuracy and efficiency were explored in~\cite{hopkins2015accuracy} for point neurons and in}~\cite{lytton2005independent} for small networks of morphologically detailed neurons.
To further improve simulation speed, a method that preserves numerical stability but allows to split \revision{single}{individual morphologically detailed} neurons across multiple parallel processes was developed in~\cite{hines2008fully}.
\revision{To reduce the complexity of individual neurons, the authors in~\cite{pozzorini2015automated} proposed a}{
A technique to convert morphologically detailed neurons into equivalent point neuron models was introuced in~\cite{pozzorini2015automated} that allows to increase simulation speed while keeping the same dynamics of the original network of detailed neurons.}

In scaling to very large networks of neurons, in silico neuroscientists soon realized the interprocess communication was becoming an important bottleneck, so simulation strategies using hybrid parallelism~\cite{plesser2007efficient}, non-blocking communication~\cite{ananthanarayanan2007anatomy}, explicit-implicit numerical schemes~\cite{kozloski2011ultrascalable} and selective sending~\cite{hines2011comparison} were introduced.
Finally, more efficient representations of the connection infrastructure~\cite{kunkel2014spiking,jordan2018extremely} and techniques for assembling the network in parallel~\cite{ippen2017constructing} proved to be essential in scaling simulation code to petascale and exascale regimes.

Despite the multiple years of research in efficient implementations of neuron models, we are still missing a more quantitative treatment of what are the actual computational characteristics of a given level of detail and how a particular level of detail may be limited by specific hardware trade-offs. How much more costly is a morphologically detailed neuron simulation compared to a representation modeling the same neuron as a point? What is the influence of how the synapses are being modeled? Can we expect that point neuron models can scale to massively parallel computers in a similar way than detailed neuron models?

\revision{We propose to use}{For the first time, we extend} performance modeling techniques \revision{}{to the field of computational neuroscience, allowing us} to establish a quantitative relationship between the parameters dictated by the biophysical model, the complexity properties of the simulation algorithm and the details of the hardware specifications.
Although we require a reasonable level of accuracy and validation against benchmarks, our goal is not to obtain highly accurate performance predictions, but rather to design a tool with sufficient generality to identify current and future bottlenecks for different levels of abstraction on the spectrum of models of neural cells' dynamics.
\revision{To achieve this goal we use analytical, grey-box performance models based}{
Based on our requirements, we choose a performance modeling approach that is neither purely based on first-principles (see e.g.~\cite{roofline:2009}) nor purely empirical(see e.g.~\cite{calotoiu_ea:2013:modeling}), but is instead an hybrid approach known as analytical, gray-box, focused} on state-of-the-art high performance computing (HPC) hardware architecture, applied to three published neural network simulations that have been selected to represent the diversity of neuron models in the literature.
Our analysis shows that features related not only to the neuron abstraction but also to the scientific question under analysis can cause variations in the hardware bottlenecks that are most influential to simulation performance.
Additionally, we find instances where the level of morphological detail is not a factor in distinguishing between models' performance, while the synaptic formalism is, \revision{breaking the myth that morphologically detailed neurons have a different performance profile than point neurons}{allowing us to identify the key factors that determine a model's computational profile}.
Finally, we show that our analysis is strongly conditioned on the published models' parameters and simulation dynamics, \revision{and}{such that} changes in some values, notably the firing frequency, can significantly alter the performance profile.

\revision{}{Our analysis demonstrates that while general-purpose computing has, until now, largely been able to deliver high performance, the next generation of brain tissue simulation will be severely limited by hardware bottlenecks.
Indeed this trend has already started,} as hardware accelerators, specifically General Purpose Graphical Processing Units (GPGPUs), have become more widespread and easier to program, and several simulators have tried to leverage the potentially large speedup promised by such devices, notably the NeMo library~\cite{fidjeland2009nemo}, the code generation framework GeNN~\cite{yavuz2016genn} and others.
We refer the interested reader to~\cite{brette2012simulating} for a review.
\revision{}{Going even further than acceleration of simulations}, intrinsic limitations in general-purpose hardware have led researchers to design custom architectures to be more similar to the brain, introducing the field of neuromorphic hardware.
One approach, used by SpiNNaker, has been to employ relatively simple, low-energy digital cores linked by a tightly connected low-latency network~\cite{sharp2011distributed}, and has been succesfully applied to a simulation of a neural microcircuit composed of 80,000 neurons and 0.3 billion synapses~\cite{van2018performance}.
A different approach was taken by the designers of the Spikey chip~\cite{pfeil2013six}, who opted for an analog representation of neural circuits able to achieve simulations faster than real time by a $10^3$ factor~\cite{wunderlich2018demonstrating}.
Although the constant improvement in computing capabilites of modern hardware has supported this growth over many years, it is likely that the rate at which this happens will soon begin to decrease, and eventually might even come to a halt~\cite{jordan2018extremely}.
Additionally, the strategy of using more and better hardware can be difficult to sustain from an economical point of view.
Therefore, we believe that the situation calls for a better, deeper understanding of how hardware capabilities interact with brain tissue simulation algorithms.
In turn, this would allow a stronger collaboration between in silico modelers, developers and hardware specialists to orchestrate the co-design of software and hardware architectures.

\revision{In conclusion, performance modeling is demonstrated to be essential in designing and optimizing future software and hardware for brain tissue simulations by providing a quantitative tool with which modelers, developers and hardware designers can collaborate.}{
In conclusion, our analysis shows that there are significant differences in the performance profiles of in silico models falling within the same category of cell-based representations.
The tools that we developed for this work constitute a quantitative tool with which modelers, developers and hardware designers can collaborate in the task of designing and optimizing future software and hardware for the next generation of brain tissue simulations.}

\subsection*{Related work}
As a testament to the growing interest of the community in the performance of brain tissue simulations, several studies on this topic have been published in the literature.
To our knowledge, however, none of them have used performance modeling as a tool to explain the empirically observed performance properties of simulations, nor have they tried to analyze such a wide scope of models as we do in this paper.

The review work of Brette et al.~\cite{brette2007simulation} presented a large number of different simulators and corresponding in silico models, but included only basic formulas for asymptotic complexity, without exploring the complicated effect that implementation and hardware have on measured performance.
In a series of papers, the developers and users of the NEST software have investigated the issues related to scaling simulations of neurons to very large scales~\cite{kunkel2014spiking,peyser2015nest,ippen2017constructing,jordan2018extremely} and proposed solutions to avoid the performance bottlenecks they encountered.
Similarly, new simulators have been implemented by Modha et al.~\cite{ananthanarayanan2007anatomy}, Kozlosky et al.~\cite{kozloski2011ultrascalable}, and new spike communication strategies have been explored~\cite{hines2011comparison} that address specific scalability issues.
All these studies provide very useful data to compare against our own model and conclusions, but are restricted to distributed simulations and focus on optimizing efficiency using novel algorithmic and implementation techniques.
A performance model for a NEST simulation was fully described in~\cite{schenck2014performance}, but focuses only on distributed simulations, neglecting single-node performance, and uses a different performance modeling approach based on interpolating empirical observations instead of the semi-analytical approach used in this work.
Focusing on very small clusters composed of only a few nodes with shared-memory capability, Hines et al.~\cite{eichner2009neural} demonstrated how to exploit multicore processor efficiently in simulations of morphologically detailed neurons, while an analysis by Gerstner et al.~\cite{zenke2014limits} concluded that real-time simulations of medium sizes plastic networks are not feasible on small clusters of CPUs, but will require accelerators or dedicated hardware.
The work on the multisplit method~\cite{hines2008fully} demonstrated that efficient acceleration of individual neurons on single compute nodes is difficult beyond a restricted number of cores, and recent work on micro-parallelism~\cite{magalhaes2019branch,magalhaes2019graph} has found significant limitations to strong-scaling due to Amdahl's law.

Several GPU implementations of brain tissue simulations have been proposed in the literature~\cite{brette2012simulating,fidjeland2009nemo,yavuz2016genn,kumbhar2019coreneuron}.
A comparison between GPUs, HPC hardware and neuromorphic hardware by Nowotny et al.~\cite{knight2018gpus} found that, under certain conditions, GPUs can beat neuromorphic hardware in terms of energy efficiency but not an HPC server in terms of performance.
However, the authors did not perform a detailed performance analysis nor used a performance model to explain this comparison, thus providing valuable yet anecdotal evidence.
Their work is tightly linked to the neuron model, the configuration of the simulation as well as the hardware being analyzed.

In the process of creating the neuromorphic hardware SpiNNaker, the designers were deeply interested by the consequences of their decisions in terms of performance.
The analysis in~\cite{painkras2013spinnaker} showed that, in the context of real-time simulations, SpiNNaker is inherently limited in its scope to a restricted subset of synaptic formalism, because the single node's memory bandwidth puts a hard limit on the number of synaptic parameters that can be streamed at every timestep.
In designing the network for SpiNNaker, a performance model was used to determine the ideal topology and network connectivity~\cite{navaridas2012analytical}, proving that performance modeling can be an extremely valuable tool in optimization and co-design for brain tissue simulations.

\subsection*{In silico models and experiments}
The approach of identifying and singling-out recurrent computational patterns within a scientific field has been applied with great success in the domain of parallel computing, leading to the definition of the dwarfs of computing~\cite{asanovic2009view}.
In computational neuroscience, the review by Brette \emph{et al.}~\cite{brette2007simulation} proposed a similar approach and introduced some fundamental concepts, such as Conductance Based (COBA) and Current Based (CUBA) formalisms for synaptic models, or point and detailed representations of neuronal morphology.
\revision{Based on}{Using the nomenclature introduced in} that review, we base our analysis of the performance landscape on three published models, chosen as representative of the extent of neuron models covered in literature.
We include an example of detailed morphology, point neuron, Conductance Based (COBA) and Current Based (CUBA) models.
We denote the set of these representative use cases as \emph{in silico models and experiments}.

\begin{figure}
    \includegraphics[width=0.9\textwidth]{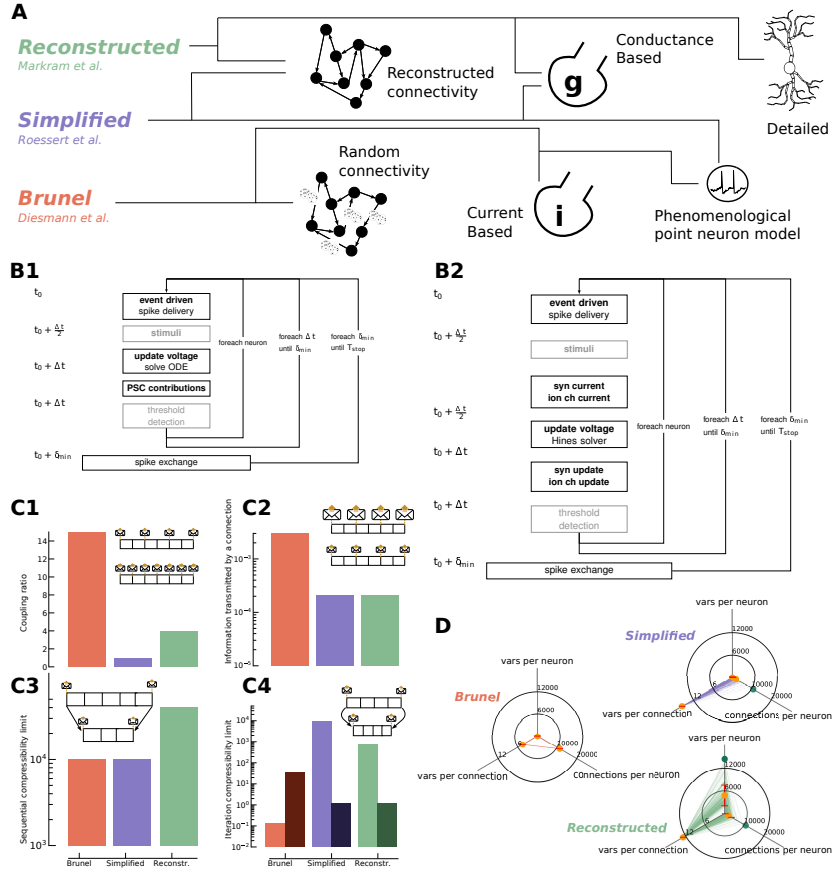}
    \caption{\bf In silico models and experiments.}
    Presentation and summary of the in silico models and experiments examined in this paper.
    \textbf{A} Main features of in silico models and experiments.
    \textbf{B1,B2} CUBA (resp. COBA) simulation algorithm.
    The simulation kernels within light grey boxes are included for completeness but are not considered in our analysis because they are not part of the computation loop.
    The larger boxes denote a synchronization point for distributed simulations.
    \textbf{C1,C2,C3,C4} Hardware-agnostic descriptive metrics allow to capture different features of the models that are relevant for performance.
    They plot, respectively, the couping ratio, the information transmitted by a connection, the sequential compressibility limit and the iteration compressibility limit.
    In the iteration compressibility limit plot, the darker bars represent event-driven computations while the light bars represent clock-driven computations.
    \textbf{D} Variability within in silico models, measured along three axis: the number of variables required to represent a neuron, the number of connections per neuron and the number of variables to represent a connection.
    \label{fig_metrics}
\end{figure}

\figurename~\ref{fig_metrics} summarizes the in silico models and experiments considered in this work.
The {\itshape Brunel} model is a randomly connected network reproducing the property of balanced excitation-inhibition that can be observed in the brain cortex.
It is based on integrate-and-fire (IAF) point neurons with CUBA dynamics~\cite{gerstner2014neuronal,brunel2000dynamics}.
As a representative example of this model, we consider here a very large-scale implementation that served as a proof of concept for the feasibility of human brain scale simulations~\cite{kunkel2014spiking}.\\
The {\itshape Reconstructed} microcircuit is based on the reconstruction of neocortical microcircuit from a mix of experimental data and first principles~\cite{markram2015reconstruction}.
This model uses morphologically detailed neurons and short-term plastic COBA synapses to capture a large amount of biological detail. \\
The {\itshape Simplified} model was obtained from~\cite{rossert2016automated} by reducing the detailed models of the Reconstructed microcircuit to pointwise GIF models~\cite{pozzorini2015automated} with similar transfer functions, while retaining the complexity from COBA synapses.

The scope of this paper is restricted to the inference phase of simulating a neural network, thus neglecting the simulation of learning, because of the additional layer of complexity that would arise from including long-term plasticity in our analysis.
Moreover, to allow for reproducibility and ease of modeling, we have switched the random synaptic release mechanism in the Reconstructed and Simplified models with a deterministic implementation of the same rule, based on an average representation.
This allows us to remove the uncertainty in the performance model due to probabilistic release, as well as the overhead from the random number generation, ultimately leading to better accuracy in the performance model.
Although synaptic noise can be considered a primary driver of cortical dynamics~\cite{nolte2018cortical}, we consider the complexity associated with random number generation in performance modeling outside of the scope of this paper.

\paragraph{Hardware-agnostic metrics to describe in silico models and experiments}
As a first approach to navigating the performance properties landscape of in silico models and experiments we propose a collection of hardware-agnostic metrics that can be computed directly from inspection of the model specifications.
These metrics have the same value regardless of the underlying simulation hardware, which can be an interesting property if one wants to compare \emph{intrinsic} model features.
To set a common ground on which we define these metrics we identify the following features shared by all in silico models and experiments: i) all neural networks can be represented by a graph where neurons are nodes and synaptic connections are edges; ii) synaptic connections can be approximated by a perfect delayed transmission of information; iii) simulation algorithms include a clock-driven portion to integrate neuronal states and an event-driven portion to integrate synaptic events (see \figurename~\ref{fig_metrics}B1,B2); iv) all simulation algorithms considered here follow the Bulk Synchronous Parallel (BSP) paradigm, where phases computation carried out independently by each parallel rank alternate with global synchronization steps, happening at fixed time intervals called minimum network delay (denoted by $\delta_{min}$).
We are thus excluding from this analysis asynchronous communication schemes such as~\cite{ananthanarayanan2007anatomy,magalhaes2019fully}, variable timestep schemes~\cite{lytton2005independent} and models that explicitly represent axons~\cite{kozloski2011ultrascalable}.

We evaluate the performance metrics of in silico models on three aspects: memory, serial complexity and information propagation.
In the memory dimension we consider aspects of the model that can affect the memory capacity footprint such as the number of parameters and degrees of freedom required to represent a neuron.
Specifically we count the number of state variables and parametes per neuron, the number of state variables and parameters per connection and the fan-in (i.e. the number of incoming connections) per neuron.
In the serial complexity dimension we consider aspects tied to sequential iterations such as timestep and the number of state variables updated at each iteration.
We define the \emph{sequential compressibility limit} as the inverse of the timestep $\Delta t$ and the \emph{iteration compressibility limit} as the number of state variables updated in a single time iteration.
Finally, in the information propagation dimension, we consider aspects tied to communication of information between neurons such as the frequency at which a global synchronization must happen and the amount of information exchanged in this step.
Here we define the \emph{coupling ratio} as the number of timesteps that can be taken before a global synchronization point must happen, given by the formula $\frac{\delta_{min}}{\Delta t}$, and the \emph{information transmitted by a connection} as the number of variables communicated by a connection on average during a mindelay period.

Each of the metrics described above can be associated to one or more performance aspects and hardware features.
For example, the low values of the coupling ratio for the Simplified and Reconstructed model (see \figurename~\ref{fig_metrics}C1) can be associated to poor strong scaling properties, while the large information transmitted by a connection (see \figurename~\ref{fig_metrics}C2) for the Brunel model translates to higher pressure on the network interconnectivity hardware.
Concerning time iterations, the large event-driven component of the iteration compressibility limit for the Brunel model (see \figurename~\ref{fig_metrics}C4) points to the fact that it could potentially be bounded by hardware latency aspects (either memory latency or critical paths in the execution) as well as dynamic imbalance, while the Simplified and Reconstructed model are more likely affected by throughput of hardware features.
The large number of connections per neuron in the Brunel model (see \figurename~\ref{fig_metrics}D) and the large number of variables to represent a neuron in the Reconstructed entail that these in silico models will be bounded by memory capacity.
Finally, the large variability of individual neurons in the Reconstructed model poses a potential static load-balancing problem, as was empirically found in~\cite{kumbhar2018performance} in the context of manycore processors.

\paragraph{\revision{Performance modeling to explore}{Using performance modeling to deliver a quantitative characterization of} the performance landscape of \revision{models}{in silico models and experiments}}

The metrics described previously can provide an insightful summary of the performance profile of in silico models and experiments, but lack the power to give quantitative performance predictions and the connection with specific hardware properties.
Therefore, we \revision{propose to}{} use performance modeling as a way of bridging the gap between biophysical models, simulation algorithms and hardware specifications.
In particular, we split the performance prediction in a single-node component and an interprocess communication component.
We address the single-node performance modeling using the Execution-Cache-Memory (ECM) model~\cite{treibig2010introducing} and the interprocess communication part using the LogGP model~\cite{alexandrov1997loggp}.
Both are well-established \revision{models}{approaches} that have been extensively validated on several hardware platforms~\cite{hager2016exploring,hager2018applying,hoefler2009loggp}, \revision{}{however significant work is required to extend and adapt them to the simulation kernels in our analysis, for example accounting for indirect memory accesses in the single-node predictions and the representation of spikes in the communication component.
For this reason,} we also present a validation of runtime predictions specific to our use-case in the Methods section.
Performance modeling enables us to explore the performance landscape of brain tissue simulations, explain current performance profiles and make predictions about future bottlenecks and trends.
Our work can provide a common framework allowing modelers, developers and hardware designers to discuss and share ideas, laying the foundations for the next generation of high performance brain tissue simulations.

\section*{Results and discussion}
Our goal is to \revision{explore}{provide a quantitative appraisal of} the performance landscape of brain tissue simulations and analyze in detail the relationship between an in silico experiment, the underlying neuron and connectivity model, the simulation algorithm and the hardware platform being used.
We carry out this analysis with the tool of performance modeling, allowing us to quantify and explain performance bottlenecks without the need of time-consuming and narrowly-scoped benchmarks.
We consider four different simulation regimes, arising from the combination of two axis.
The first axis is defined by the strategy for parallelization: shared-memory or distributed-memory.
The second axis is defined by the strategy for scaling the problem size with the available hardware: maximum-filling or real-time, corresponding respectively to weak-scaling or strong-scaling.
Given that the results of this analysis are tightly dependent on the hardware platform under consideration, the reference single-node architecture is Intel's Skylake-X processor with AVX512 vectorization, considered to be a prototypical example of state-of-the-art HPC microarchitectures.
For distributed simulations, the reference network architecture is a vendor-optimized HPE implementation of the MPI standard over an Infiniband EDR 100 GB/s fabric.
\subsection*{Shared-memory maximum filling}
One of the most common simulation configurations involves scaling the number of neurons until the memory capacity limit is reached.
This configuration has been used as proof-of-concept for brain tissue simulations to the scale of brain regions~\cite{ananthanarayanan2009cat,jordan2018extremely} and even the full brain~\cite{izhikevich2008large}, and constitutes a fundamental tool for neuroscientists to simulate networks whose sizes are representative of the neural systems they are studying.

\begin{figure}
    \includegraphics[width=0.99\textwidth]{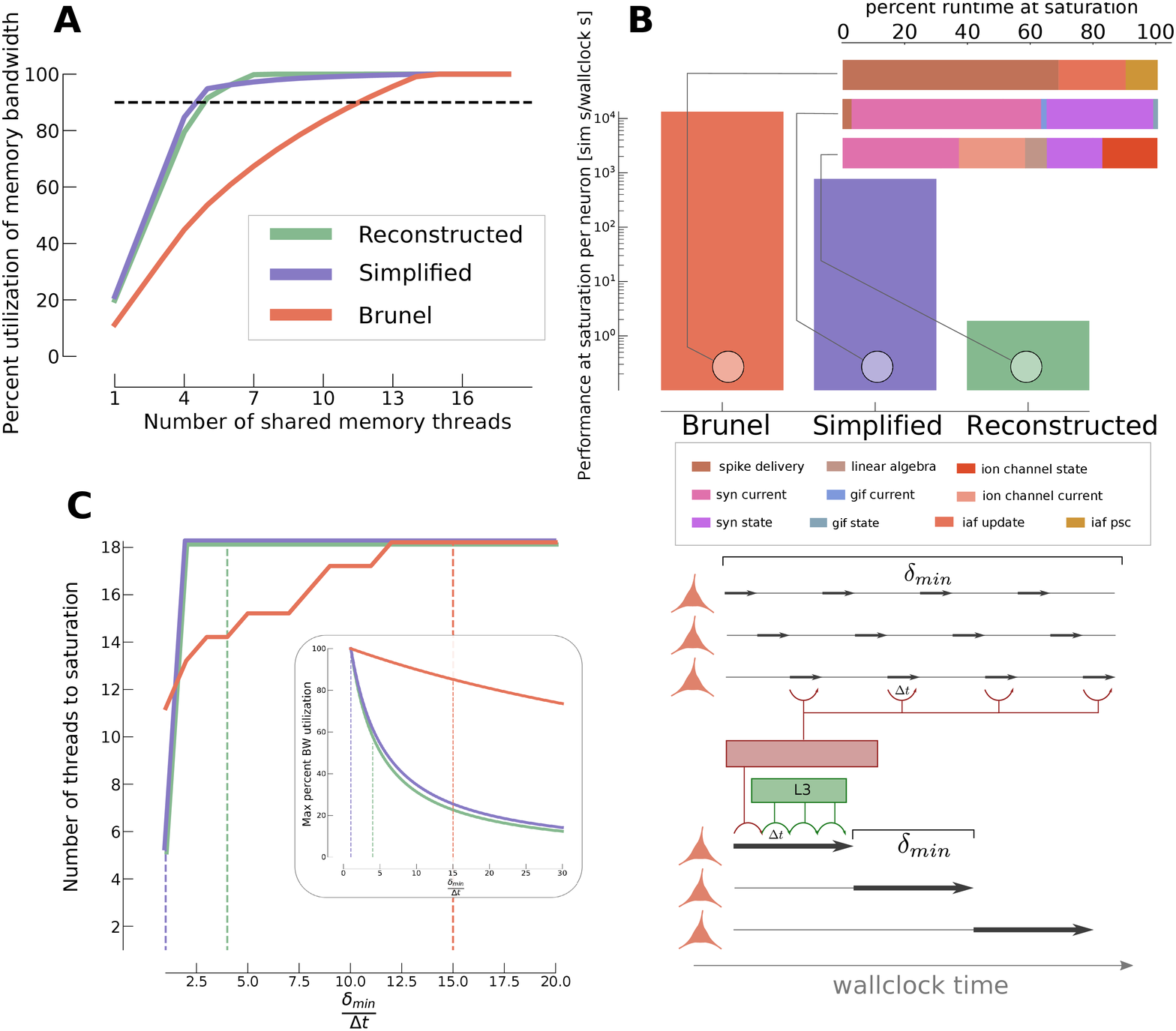}
    \caption{\bf Memory bandwidth saturation of in silico neuron models on a state-of-the-art HPC architecture.}
    The memory interface utilization and total runtime of different neuron models, as predicted by the ECM performance model on a Skylake-X architecture with AVX512 vectorization.
    \textbf{A} Percentage of memory bandwidth utilization for different neuron models, as a function of the number of shared memory threads.
    The dashed black line denotes the threshold for saturation set at 90\% utilization.
    \textbf{B} Simulation performance per neuron in simulated seconds per wallclock second under conditions of full saturation of the memory bandwdith.
    To obtain the performance prediction for a whole network divide the performance values by the number of neurons in the network.
    Note the logarithmic scale on the y axis.
    The {\itshape inset} shows the percent of the total runtime required by each kernel within a model's simulation loop.
    \textbf{C} To mitigate the effect of memory bandwidth saturation, a smart ordering of time and neuron loops is implemented by state-of-the-art simulators, as shown in the diagram on the right.
    We plot the number of threads required to reach 90\% saturation of memory bandwidth as a function of the coupling ratio, clipping the total to the maximum allowed by the reference architecture.
    The inset shows the maximum bandwidth utilization as a percent of the theoretical peak bandiwdth, as a function of the coupling ratio.
    \label{fig_nsatur}
\end{figure}

\paragraph{Memory bandwidth limits shared-memory parallelism}
Modern architectures are typically designed with memory bandwidth as the most relevant bottleneck for shared-memory parallelism~\cite{McCalpin:1995}.
This means that if all the shared memory parallel threads are used, it is very likely that performance will be bounded by the memory bandwidth.
Indeed, this has been demonstrated to be the case for simulations of detailed neurons~\cite{cremonesi2019analytic} and strongly suspected in the case of point neurons~\cite{zenke2014limits}.
To verify the hypothesis that memory bandwidth could indeed be the main bottleneck in the maximum filling regime we \revision{used}{extend} the ECM performance model \revision{}{following our previous work~\cite{cremonesi2019analytic}} to predict the utilization of the memory interface, computed as a fraction of the theoretical maximum memory bandwidth, for the three in silico models considered in this work.
The results are shown in \figurename~\ref{fig_nsatur}A.
we \revision{found}{find} that all models pass the threshold of 90\% utilization well before all available parallel threads are utilized, meaning that memory bandwidth is indeed a bottleneck in the maximum filling scenario, under the assumption that data must be pulled from main memory \emph{at every time iteration}.
Moreover, this also implies that the parallelism exposed by the architecture cannot be fully exploited in the maximum filling regime.
However, we surprisingly also \revision{found}{find} that, regardless of the level of morphological detail, COBA models share a similar pattern of early saturation, while the CUBA IAF model requires more parallelism to achieve memory bandwidth saturation.

\paragraph{State-of-the-art HPC memory chips can sustain fast simulations of the Brunel and Simplified models at full saturation}
From a practical point of view, in addition to analyzing the scaling behaviour of simulations, computational neuroscientists are also interested in predicting the actual runtime for a given model.
Therefore, we predict\revision{ed}{} the simulation performance for the three in silico models under the assumption that \emph{memory bandwidth is fully saturated}.
The results are plotted in \figurename~\ref{fig_nsatur}B, and Table~\ref{tab_mem} reports the memory traffic requirements to simulate one second of activity, alongside the predicted memory capacity.
As unit of measure for performance we chose simulated seconds per wallclock second per neuron, in order to present our results in a way that is independent from the network size and the duration of the simulation.
Our results indicate that the modern, fast memory chips on the reference architecture are able to sustain faster-than-real-time simulations of up to roughly $10^4$ neurons in the Brunel model, and $10^3$ neurons in the Simplified model, while faster-than-real-time simulations of the Reconstructed model are predicted to be impossible.
As an important remark, these predictions only consider the \emph{computational and communication kernels} of an in silico model, and notably neglect event bookkeeping efforts, random numbers generation and computation of stimuli.
In a real world scenario, the empirically measured performance of a model could be much smaller if the model's execution is not dominated by computational aspects.

\begin{table}[!ht]
\centering
\caption{
{\bf Predicted memory traffic and capacity requirements.}}
\begin{tabular}{|l|l|l|}
\hline
     & {\bf Volume per biological millisecond [kB]} & {\bf Capacity [kB]} \\ \thickhline
Brunel        & $8$                 & $4.5\times 10^3$ \\ \hline
Simplified    & $1.3\times 10^2$    & $9.9\times 10^2$ \\ \hline
Reconstructed & $5.9\times 10^4$    & $3.9\times 10^4$ \\ \hline
\end{tabular}
\begin{flushleft}
    The requirements in this Table are computed considering only the data structures strictly relevant to computation, thus neglecting overhead from implementation details.
\end{flushleft}
\label{tab_mem}
\end{table}

\paragraph{Event-driven synaptic integration dominates CUBA performance, while clock-driven kernels dominate COBA performance.}
We also predict\revision{ed}{} the breakdown of relative importance of different kernels that constitute the simulation algorithm of a model, shown in the inset of \figurename~\ref{fig_nsatur}B, allowing us to highlight some interesting differences between models.
In the maximum filling scenario\revision{, during a purely shared-memory execution,}{} the Reconstructed model is not dominated by a single kernel.
Instead, synaptic and ionic current kernels constitute almost 60\% of the execution time, while state update kernels, which are commonly regarded as more costly, are a few points short of taking up the remaining 40\%.
This has been extensively validated in~\cite{cremonesi2019analytic} and can be explained by the fact that current kernels have stronger data requirements and lower computational requirements, and thus poorer performance when the bottleneck is constituted by the memory bandwidth.
The Simplified model is similarly dominated by the computation of synaptic current kernels.
Since the Simplified and the Reconstructed model share the same COBA synaptic formalism, this can explain why some of their performance properties are very similar, in spite of the fact that their representation of neurons' morphologies is extremely different.
Conversely, the performance of the Brunel model is determined for more that 60\% by a single kernel: the event-driven integration of synaptic events.
In the following sections we explore in detail how this characteristic has an impact on determining which hardware feature is most relevant for the performance of in silico models.

\paragraph{Ordering of loops to avoid memory bandwidth saturation.}
State-of-the-art simulators employ a specific ordering of the loops over neurons, timesteps ($\Delta t$) and minimum network delay steps ($\delta_{min}$) to minimize the impact of memory bandwidth by maximizing data locality.
This strategy is exemplified on the right of \figurename~\ref{fig_nsatur}C, and consists of allowing a neuron to step in time from the beginning of the minimum network delay interval to its end, before initiating the stepping of the second neuron.
This approach is equivalent to swapping the \emph{neuron} and $\Delta t$ loops in \figurename~\ref{fig_metrics}B1,B2, and has the benefit of maximizing data locality.
Depending on the memory traffic requirements, data could be stored in the cache after being read once from main memory during the first timestep, ready to be used for the next time iterations within the same minimum network delay period.
Throughout this work we make the conservative assumption that, when using this strategy, data must be fetched from main memory on the first timestep and from the L3 cache on consecutive timesteps.
In special cases where the memory traffic requirements of a single neuron are very low and the number of synaptic events integrated in a timestep is sufficiently small, data could potentially come instead from higher levels in the hierarchy such as the L2 cache; however for the sake of simplicity we make the assumption that reused data must always be fetched from L3.
The number of timesteps within a minimum delay period has of course a great influence on the effectiveness of this strategy in terms of reducing pressure on the memory bandwidth.
To \revision{measure}{quantify} this, we predict\revision{ed}{} the maximum memory bandwidth utilization as a function of the coupling ratio defined by $\frac{\delta_{min}}{\Delta t}$.
For COBA models this follows a steep decline for small values of the coupling ratio, indicating that even just a few timesteps exploiting data locality can provide great benefits in terms of shared memory scaling, while in the case of the CUBA model this decline is much slower, which could explain why the published Brunel model~\cite{kunkel2014spiking} has such a large value of the coupling ratio while the Reconstructed and Simplified model have low values (see \figurename~\ref{fig_metrics}C1).

\subsection*{Distributed maximum filling}
To overcome the limit on network size imposed by the memory capacity of individual compute nodes, computational neuroscientists have begun executing parallel distributed simulations on a cluster of compute nodes.
In the distributed maximum filling regime we still consider that the memory capacity limit of each individual compute node will be maxed out, but we introduce the possibility of increasing the number of interconnected compute nodes as a means of increasing the computational power and capacity of the architecture.
In terms of scaling the problem size proportionally to the number of compute nodes, the fact that we assume the memory capacity limit to be always maxed out translates to the concept of weak scaling, on which we will base our analysis in this section.
To keep our initial analysis simple we ignore the loop-ordering optimization described above and assume instead that the memory bandwdith of individual compute nodes will always be saturated.
We will lift this assumption in later sections.

\begin{figure}
    \includegraphics[width=0.99\textwidth]{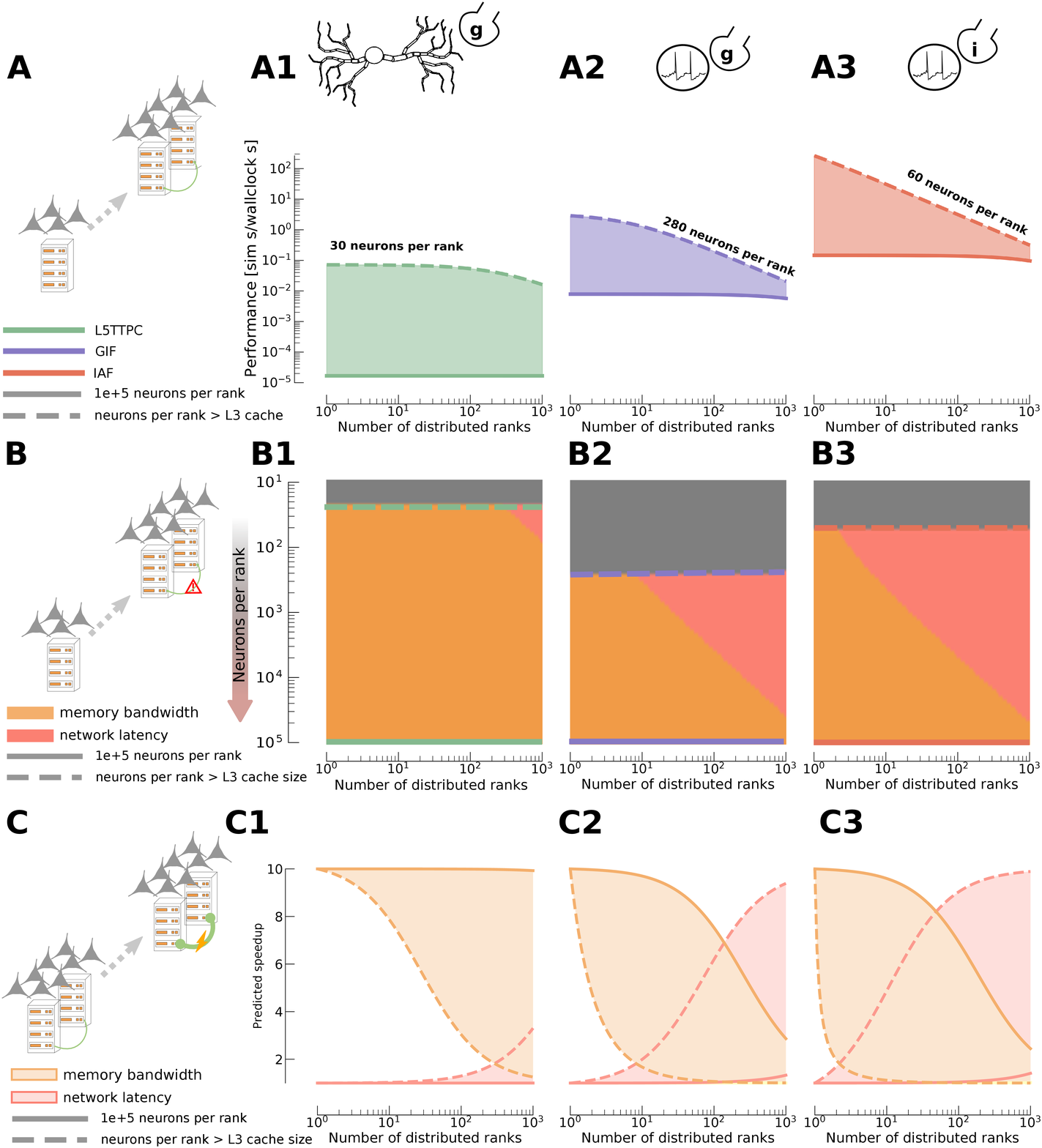}
    \caption{\bf Performance of distributed max-filling scenario and most relevant hardware bottlenecks.}
    \textbf{A):} predicted performance of the three \emph{in silico models} in a weak-scaling, max-filling scenario.
    We plot several values of the number of neurons per rank, ranging from the smallest number that would exceed a typical L3 size of 25\MB~to $10^5$ neurons per rank.
    The predicted performance is plotted as the number of model seconds for which the whole network can be simulated in 1 wallclock second.
    We make the assumption of full memory-bandwidth saturation.
    We take the SKX-AVX512 node architecture connected by an Infiniband EDR 100 \GBS~network as representative of an HPC cluster.
    \textbf{A1,A2,A3):} represent the detailed COBA, simplified COBA and point CUBA models, respectively.
    \textbf{B):} hardware bottlenecks predicted by the performance model as a function of the number of neurons per rank (y-axis) and the number of distributed ranks (x-axis).
    The grey regions denote an \revision{{\itshape impossible}}{} area \revision{in which}{where} the number of neurons per rank would require less memory than a typical L3 cache size, thus invalidating the saturation assumption.
    \textbf{B1,B2,B3):} represent the detailed COBA, simplified COBA and point CUBA models, respectively.
    \textbf{C):} Speedup predicted by the model when a single hardware feature is improved by a factor of 10x, as compared to the baseline performance using the HPC architecture described in \textbf{A}.
    Again, since the number of neurons per rank plays an important role, the shaded areas represent the speedups for different values of the number of neurons per rank, in the range described in \textbf{A}.
    \textbf{C1,C2,C3):} represent the detailed COBA, simplified COBA and point CUBA models, respectively.
    \label{fig_max_fill}
\end{figure}

\paragraph{Weak scaling properties and performance predictions of in silico models}
We predict\revision{ed}{} the performance of distributed simulations in a weak scaling, maximum filling scenario for different numbers of neurons per rank and all in silico models, using our performance model.
Results are presented in \figurename~\ref{fig_max_fill}A, where the solid lines correspond to $10^5$ neurons per rank while the dashed lines correspond to a small number of neurons per rank, computed such that its memory footprint barely exceeds the 25\MB~of the reference architecture's L3 cache; smaller values would not be possible because they would break the memory bandwidth saturation assumption.
As expected, for a fixed configuration and a small cluster size the Brunel model has the best predicted performance, beating by roughly a factor 10 the peformance of the Simplified model and roughly a factor $10^4$ the performance of the Reconstructed model.
Interestingly, these differences are much less pronounced for large cluster sizes, where the difference between the Brunel and Reconstructed model can be reduced to less than a factor 10.
Networks with the largest numbers of neurons considered here possess excellent weak scaling properties regardless of the underlying modeling abstraction, at least up to the cluster sizes considered here.
Conversely, only small networks of the Reconstructed model retain the same quasi-optimal scaling behaviour, while both the Brunel and Simplified model suffer from performance degradation at large cluster sizes, with the average rate of degradation being significantly larger for the Brunel model.

\paragraph{Explaining weak scaling performance properties through bottleneck analysis}
\figurename~\ref{fig_max_fill}B explains the degradation of performance by identifying the most relevant bottlenecks for different configurations.
Here a hardware bottleneck is defined as the most relevant hardware feature as computed by our performance model, as explained in the Methods Section.
The first striking property revealed by this analysis is that network bandwidth is never a bottleneck for the in silico models considered here.
Instead, large-scale simulations are dominated by the latency of the collective communication.
This points to the fact that investigating spike communication strategies such as neighborhood collectives~\cite{jordan2018extremely}, nonblocking point-to-point schemes~\cite{ananthanarayanan2007anatomy} or asynchronous execution~\cite{magalhaes2019graph} is essential to reach brain-scale simulations.

\paragraph{Distributed simulations of large networks can be improved by increasing memory bandwidth, small networks by decreasing network latency}
The aforementioned bottleneck analysis is useful to understand which is the most important hardware feature in a given simulation configuration, but it does not provide any information about the relative importance of the other features.
Therefore we \revision{used the performance model to}{}predict the speedup corresponding to a $10\times$ improvement of a single hardware feature, for different numbers of neurons per distributed rank, as a function of the number of ranks.
The results are plotted in \figurename~\ref{fig_max_fill}C, which shows that at small cluster sizes all models would benefit from an improvement in the nodes' memory bandwidth, but not from an improvement in network latency, while the situation is reversed at large cluster sizes.
The predicted speedup obtained by improving the memory bandwidth degrades much faster for simulations with a small number of neurons per rank versus a large number of neurons per rank, while the opposite is true for the speedup coming from an improved network latency.

\subsection*{Single-node simulations in real time}
Another widespread simulation regime is not focused on simulating the largest possible network, but in simulating a fixed size network as fast as possible.
We call this the real time regime, because that represents an ideal target for the performance of a simulation.
Currently, on the one hand it is unclear whether real time is realistically achievable on modern hardware~\cite{zenke2014limits}, and on the other hand special hardware that breaks this limit by design has already been conceived and tested for small networks~\cite{aamir2018accelerated}.

\paragraph{Memory bandwidth dominates the shared-memory strong scaling of Brunel and Simplified models, while a mix of hardware features influences the performance of the Reconstructed model}
Given that it is possible to observe superlinear speedup as the dataset is made increasingly small by virtue of it fitting into faster cache memory, we predict the performance per neuron of all in silico models assuming the dataset could be fully contained in different levels of the memory hierarchy.
For simplicity, we neglect\revision{ed}{} the fact that some of these model and cache combinations are infeasible in practice, e.g., due to the memory footprint of a single neuron in the Reconstructed model exceeding the L1 cache size.
For DRAM, we assume that the ordering of loops technique is used to avoid saturation of the memory bandwidth, thus extending our analysis by discarding the saturation hypothesis made in previous sections.
For all performance predictions we assume that all available threads in the reference architecture ($18$ in total) are being utilized.
Results are reported in Table~\ref{tab_real_time}, where we give the raw performance value when data is in DRAM, and the corresponding (superlinear) speedup factor as the dataset becomes small enough to fit in higher levels of the cache hierarchy.

\begin{table}[ht!]
\begin{adjustwidth}{-0.25in}{0in} 
\centering
\caption{
{\bf Predicted performance per neuron without memory bandwidth saturation assumption, considering all available threads are used.}}
\label{tab_real_time}
\begin{tabular}{|l|c|c|c|c|}
    \hline
     model & performance (DRAM) & speedup in L3 & speedup in L2 & speedup in L1 \\ \thickhline
     Brunel & $3.9 \times 10^4$ & 2.5 & 8.7 & 33.9 \\ \hline
     Simplified & $7.9 \times 10^2$ &  4.7 & 6.5 & 6.6 \\ \hline
     Reconstructed & 3.9 & 1.8 & 2.4 & 2.4 \\ \hline
\end{tabular}
\begin{flushleft}
Performance is measured in simulated seconds per wallclock second.
\end{flushleft}
\end{adjustwidth}
\end{table}

\begin{figure}
    \includegraphics[width=0.99\textwidth]{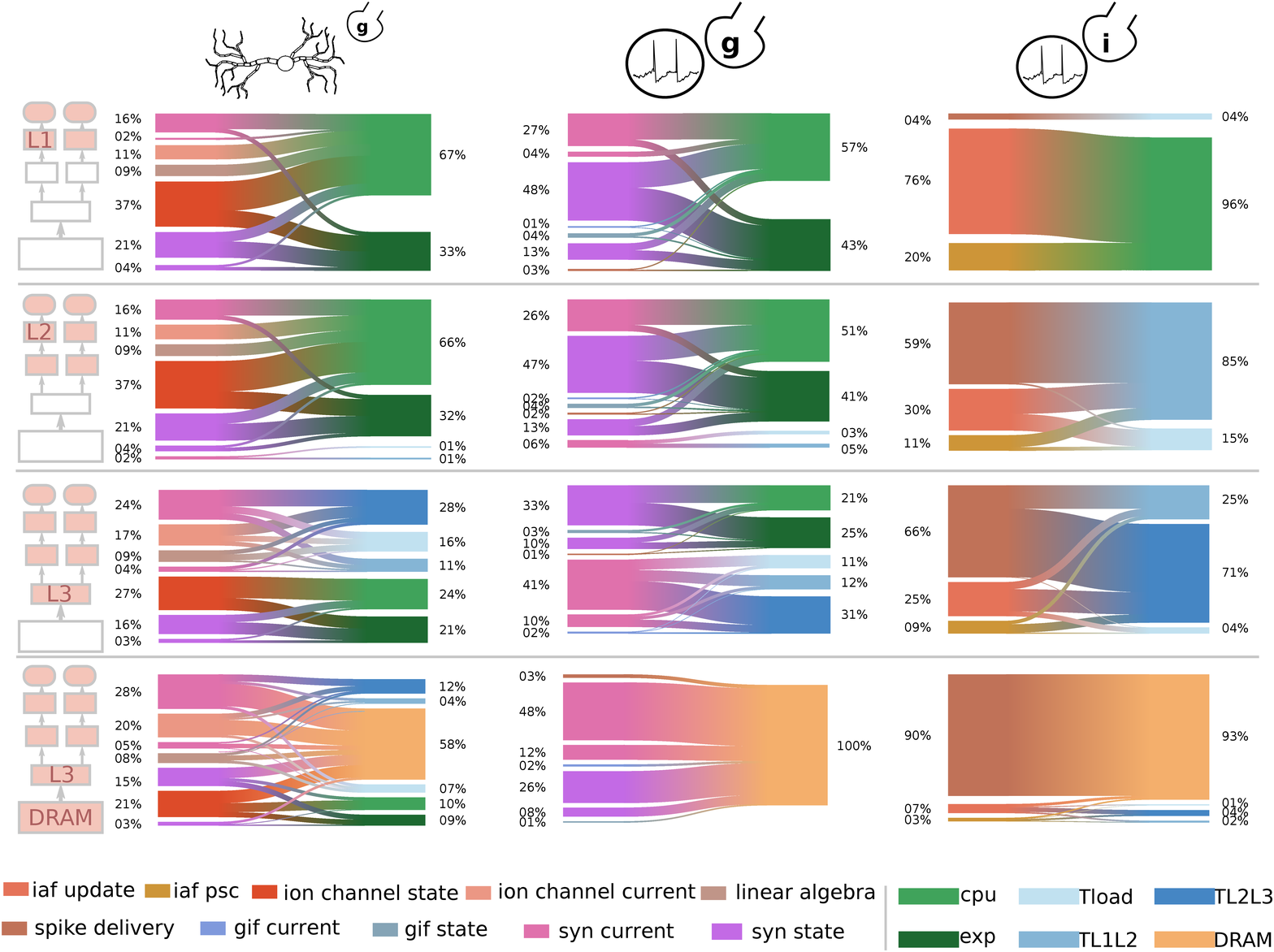}
    \caption{\bf Overall runtime contributions from algorithmic kernels and hardware features.}
    For each level of the cache hierarchy we show the relative contribution of algorithmic kernels to the total runtime, and the corresponding breakdown into contributions from hardware features.
    We assume a single node of the SKX architecture with AVX512 vectorization and using hte maximum number of threads.
    The bar plots show the predicted speedup relative to DRAM performance when the number of neurons is computed to fill exactly the corresponding cache level.
    For DRAM, the table presents the performance per neuron measured in [simulated s / wallclock s], assuming the loop ordering technique to avoid saturation is used.
    \label{fig_hw_contrib}
\end{figure}

\figurename~\ref{fig_hw_contrib} shows the predicted performance breakdown into simulation kernels as well as hardware features for all in silico models, assuming that the dataset fits in different levels of the memory hierarchy.
When data is in the highest level of the cache hierarchy (L1), the most important kernels for all models are state update kernels, and the most relevant hardware feature is the CPU throughput.
Additionally, in the COBA models the computation of the exponential (for updating the synaptic states) constitutes a signficant portion of the overall execution time.
As the dataset increases in size and is only able to fit in lower levels of the cache (L2 or L3) the predicted performance of the Brunel model rapidly degrades, while the COBA models' remains more stable.
In practice, this is an indication that the Brunel model is bounded by the data path while the COBA models are bounded by the maximum achievable FLOPs.
Our breakdown analysis confirms this, although for the reference architecture COBA models are best represented by a mix of core-bound and data-bound kernels, especially when the dataset fits only in the L3 cache.
Complementarily, in COBA models the relative importance of the core-bound state update kernels gradually loses weight in favour of data-bound current kernels, while in the Brunel model the weight of the spike delivery kernel gradually increases, eventually becoming the most relevant kernel in the execution.
In spite of this technique, both point neuron models are clearly dominated by the saturation of the memory bandwidth.
\revision{}{In particular, the fact that memory bandwidth is the only factor in determining the performance of the Simplified model can be directly related to the fact that its coupling ratio has a value of 1, as shown in~\figurename~\ref{fig_metrics}.}
The performance profile of the Reconstructed model is more diverse, and while 60\% of the execution time is still dominated by memory bandwidth, the data transfers between the caches, arithmetic instructions, and throughput of exponential function evaluations also take up a significant portion of the runtime.
Again we stress that this phenomenon is tightly linked to the hardware architecture used as reference, specifically to its balance of compute power, memory bandwidth and number of threads.
However, we can assume that the general application of these conclusions still holds because most modern architectures are designed with a similar balance point~\cite{McCalpin:1995}.

\subsection*{Distributed real time}
An effective strategy for improving simulation performance or to handle larger networks is to dedicate more hardware to the task, distributing the simulated neural network across multiple compute nodes.
For a fixed problem size, this translates to the concept of strong scaling.
The limits to strong scaling of medium sized plasticity networks have been empirically explored in~\cite{zenke2014limits}.
Here we generalize their analysis to other in silico models as well as provide clear explanations for the causality of bottlenecks, backed by our performance model.

\begin{figure}
    \includegraphics[width=0.99\textwidth]{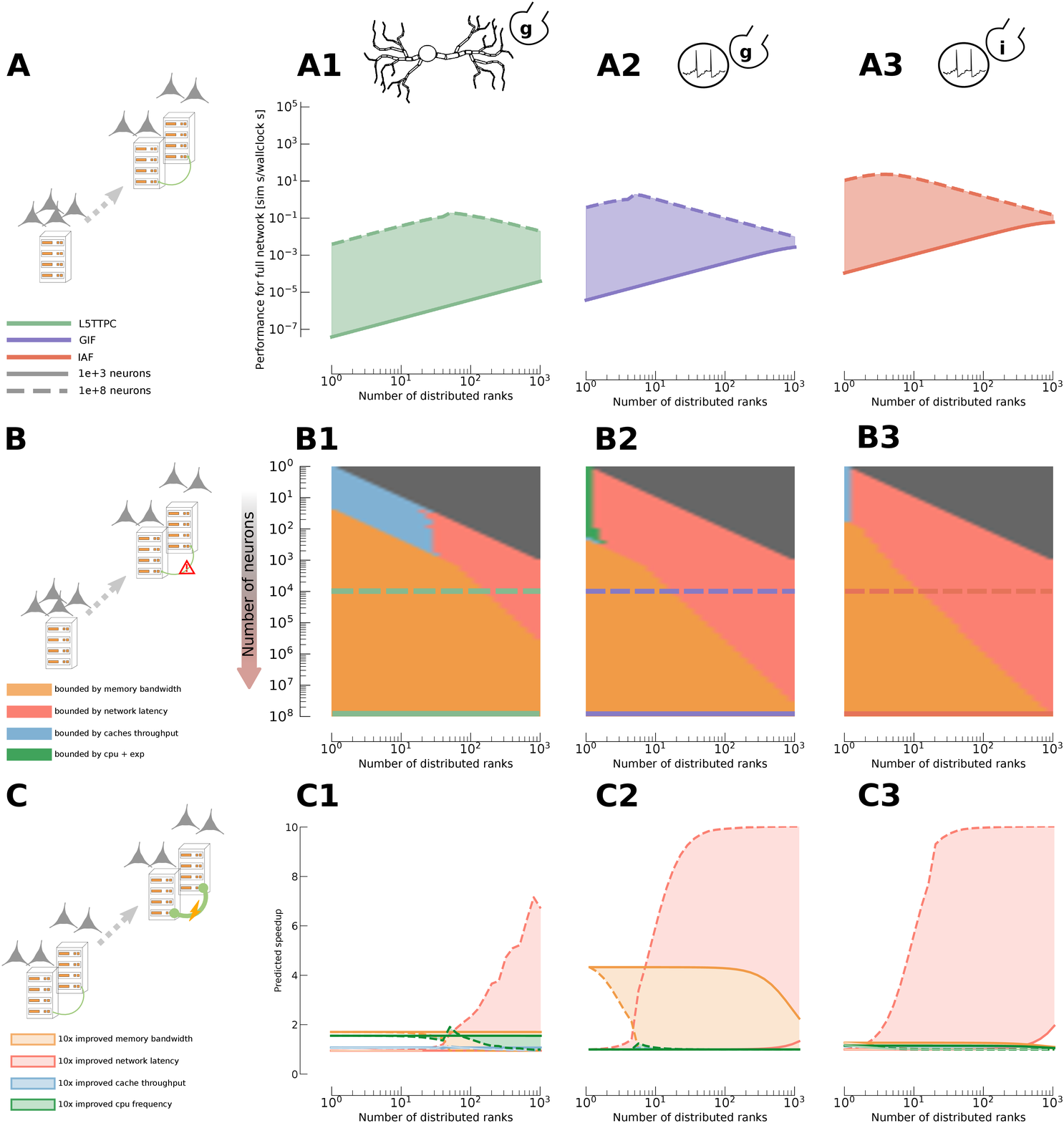}
    \caption{\bf Performance of distributed real time strong-scaling scenario and most relevant hardware bottlenecks.}
    \textbf{A):} predicted performance of the three \emph{in silico models} in a strong-scaling scenario, as a function of the number of distributed parallel processes.
    We restrict our analysis to clusters of up to $10^4$ distributed parallel processes, based on the SKX-AVX512 architecture connected by an Infiniband EDR 100 GB/s network.
    Since the total number of neurons can be an important factor for performance, we plot performance for different network sizes.
    The dashed and solid lines represent the performance of simulations with networks of $10^4$ and $10^8$ neurons respectively.
    We plot the predicted performance measured as the number of model seconds for which the whole network can be simulated in 1 wallclock second.
    Contrary to the max-filling scenario, we do not impose the hypothesis of memory bandwidth saturation, although we still assume that, whenever possible, all shared memory threads available will be used.
    \textbf{A1,A2,A3):} represent the detailed COBA, simplified COBA and point CUBA models, respectively.
    \textbf{B):} hardware bottlenecks predicted by the performance model as a function of the number of neurons in the whole network (y-axis) and the number of distributed ranks (x-axis).
    The grey areas denote an impossible configuration that would require splitting of individual neurons.
    \textbf{B1,B2,B3):} represent the detailed COBA, simplified COBA and point CUBA models, respectively.
    \textbf{C):} Speedup predicted by the model when a single hardware feature is improved by a factor of 10x, as compared to the baseline performance using the reference HPC architecture.
    The dashed and solid lines have the same meaning as in \textbf{A}.
    \textbf{C1,C2,C3):} represent the detailed COBA, simplified COBA and point CUBA models, respectively.
    \label{fig_real_time}
\end{figure}

\paragraph{Performance predictions for strong scaling of arbitrarily sized networks}
We predict\revision{ed}{} the performance and scaling bottlenecks of in silico models in a strong scaling scenario, using our performance model, and present the results in \figurename~\ref{fig_real_time}.
As in the distributed maximum filling scenario, the problem size has a major impact on performance, so we \revision{decided to }{}include several possible sizes in our analysis.
Some researchers have explored the possibility of splitting a neuron across more than a single parallel process, such as the multisplit method~\cite{hines2008fully}, branch-parallelism~\cite{magalhaes2019branch} and domain decomposition~\cite{kozloski2011ultrascalable}.
We do not include such techniques in our analysis because their granularity falls outside the scope of this work.
In practice, this imposes a limit on how many distributed ranks a given neural network could be scaled on.
In \figurename~\ref{fig_real_time}A we present raw performance predictions for different network sizes in a strong scaling scenario.
The dashed lines correspond to the minimum network size such that strong scaling can be carried out until occupancy of the full cluster (set here at $10^4$ distributed ranks), while thinner solid lines represent smaller network sizes that cannot be scaled to the full cluster.
For all in silico models, as long as the network size is sufficiently large, the performance initially improves as we distribute the problem over increasingly more ranks.
However, for all in silico models there exists a threshold number of ranks after which the benefits from adding hardware become less prominent.
Scaling to larger cluster sizes after this threshold can be counter-productive, and even result in performance degradation.
The threshold value itself is a function of the hardware architecture, in silico model and problem size.
Interestingly the striking differences in performance between in silico models at small cluster sizes can be evened out quite significantly at large cluster sizes in this scenario.
For example, simulating a large Brunel network on 10 distributed ranks can be roughly four orders of magnitude faster than a Reconstructed network on the same hardware, but the difference between models goes down to two orders of magnitude at large cluster sizes.

\paragraph{Network latency and memory bandwidth are the main bottlenecks in strong scaling}
Following the same procedure of the maximum filling scenario we investigate the reasons for performance degradation by plotting the most signficant hardware bottlenecks for all combinations of network size and cluster size in \figurename~\ref{fig_real_time}B.
We assume that the time and neuron loops are ordered to minimize pressure on the memory bandwidth.
Even though we do not make the explicit assumption of memory bandwidth saturation, this hardware feature is still among the most relevant for all in silico models (as was also shown in \figurename~\ref{fig_hw_contrib}).
Moreover, network bandwidth is never the dominating bottleneck for all in silico models and all configurations, while network latency always becomes the most important bottleneck at large cluster sizes.
At very small cluster sizes, we recover the results from \figurename~\ref{fig_hw_contrib}.

\paragraph{Network latency gives significant performance improvements for point neuron models, but no improvement in a single factor would be sufficient to increase performance in the Reconstructed model}
To further investigate the relevant bottlenecks, \revision{using our performance model we predicted}{we predict} the expected speedup corresponding to a tenfold increase in a single hardware feature and present the results in \figurename~\ref{fig_real_time}C.
Both point neuron models show a similar structure: an improvement in the features of a single node such as CPU frequency or cache throughput yields an improvement in performance only for very small networks, while a tenfold improvement in network latency would guarantee a significant improvement in performance for networks of all sizes and sufficiently large cluster sizes.
The only difference between the two point neuron models in this analysis \revision{is the impact of}{are the benefits to be gained from improving the} memory bandwidth: while \revision{it has almost no impact on}{there would be almost no benefit for} the Brunel CUBA model, it has a moderate influence in the performance of the Simplified COBA model, especially for large networks.
The situation for the Reconstructed model differs, and there is no single factor that would result in a significant performance improvement at any network size, except for strong scaling of very small networks where the CPU throughput is the dominating hardware feature.
This can be explained by the diversity of relevant hardware factors identified in \figurename~\ref{fig_hw_contrib}: when a single hardware feature is improved, another bottleneck is quickly reached and the total resulting performance improvement is suboptimal.
For large cluster sizes the situation normalizes to that of the other in silico models, and network latency becomes the dominant factor, such that a tenfold increase results in a nearly-equivalent performance boost, especially for small networks.

\section*{Dependence of performance on model parameters}
Parameters of the in silico models have an important, yet often difficult to explain, impact on performance.
We test\revision{ed}{} the impact of the firing frequency, minimum network delay and fan-in using our performance model, and present the results in this section.
We neglect\revision{ed}{} the timestep because it has a generally straightforward relationship with performance, and was omitted for brevity.


\begin{figure}
    \includegraphics[width=0.95\textwidth]{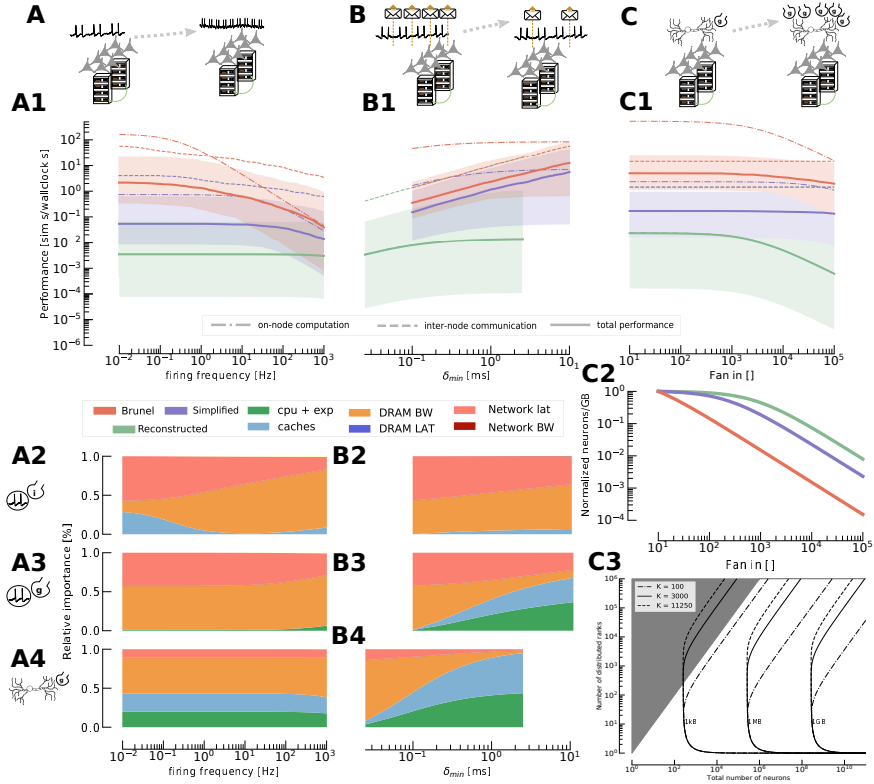}
    \caption{\bf Effect of model parameters on performance.}
    \textbf{A} Effect of firing frequency on the performance of distributed simulations.
    \textbf{A1} Median performance (total and broken down into communication and on-node computation components) as a function of firing frequency.
    The median is computed over one thousand randomly generated samples of number of neurons and number of distributed ranks.
    The shaded area surrounding the total performance represents the 25th and 75th percentiles.
    \textbf{A2,A3,A4} Stacked plot of the median relative contributions from hardware features as a function of firing frequency, respectively for the Brunel, Simplified and Reconstructed model.
    The median is computed over one thousand randomly generated samples of number of neurons and number of distributed ranks as above.
    \textbf{B} Effect of $\delta_{min}$ on the performance of distributed simulations.
    \textbf{B1} Median performance as a function of $\delta_{min}$.
    The distribution for computing the median and the shaded areas are the same as in A1.
    \textbf{B2,B3,B4} Stacked plot of the median relative contributions from hardware features as a function of $\delta_{min}$, respectively for the Brunel, Simplified and Reconstructed model.
    The range of acceptable values for $\delta_{min}$ changes across different in silico models because these values were computed as multiples of the model's timestep, i.e. by using the definition of coupling ratio.
    \textbf{C} Effect of fan-in the performance of distributed simulations.
    \textbf{C1} Median performance (total and broken down into communication and on-node computation components) as a function of fan-in.
    Fan-in has a direct effect on the memory requirements of an in silico model, which in turn is potentially an important factor for performance because it affects the maximum allowed number of neurons per rank.
    For this plot, we assumed that all randomly generated configurations were possible, i.e. that memory capacity was not an issue.
    \textbf{C2} Contour plot of predicted memory requirements of the connections table, as a function of the total number of neurons (x-axis) and number of distributed ranks (y-axis).
    The contour levels corresponding to 1\KB, 1\MB~and 1\GB~are shown for different values of the fan-in.
    \textbf{C3} Memory requirements as a function of the fan-in, normalized by the memory requirements of a model with 10 incoming synapses.
    \label{fig_parameters}
\end{figure}

\paragraph{Firing frequency's differential effect on communication and computation}
Firing frequency is commonly cited as one of the most impactful parameters on simulation performance~\cite{yavuz2016genn}.
In this work we consider it a parameter even though it usually cannot be explicitly set by the user, and is instead an emerging property of the simulation.
\figurename~\ref{fig_parameters}A shows the predictions of our performance model for the three in silico models, for values of the firing frequency in a physiological range.
To take into account the obvious fact that the total number of neurons and distributed ranks can introduce a large variability in our predictions, we randomly generate\revision{d}{} 1000 value pairs of [number of neurons, number of distributed ranks] and plot the median predicted performance, additionally broken down into its two components of inter-node communication and on-node computation.
Firing frequency has an effect on communication by changing the size of the spike message as well as on computation by changing the amount of events that must be integrated by neurons.
In \figurename~\ref{fig_parameters}A1 it is noticeable that there exists a threshold frequency below which $f$ does not affect performance, but once this threshold is passed firing frequency becomes a primary factor, inducing a linear, almost unit-slope degradation in performance.
This effect is clearly visible in the Simplified model, and even more so in the Brunel models.
For the Reconstructed model this threshold value exists, but is so large that we can safely assume that, in the median case, firing frequency has no effect on performance.
To investigate the reasons for performance degradation we look\revision{ed}{} at the breakdown of relative importance of different hardware features as a function of the firing frequency, plotted in \figurename~\ref{fig_parameters}A1,A2,A3.
Similarly to before, we randomly generate\revision{d}{} 1000 couples of [number of neurons, number of distributed ranks] but we plot the mean relative importance instead of the median to keep the total constantly equal to 100\%.
Our analysis shows an interesting behaviour: as the firing frequency becomes larger, the relative pressure on the memory bandwidth (and eventually the network bandwidth) becomes larger, while the relative pressure on the network latency becomes smaller.
So not only there is more computation to be done as firing frequency gets larger, but also the mix of hardware bottlenecks changes.
This behaviour was observed empirically not only on general-purpose CPUs~\cite{zenke2014limits} but also on GPUs \revision{}{which are additionally more susceptible to dynamic load balancing because of the extrememly large number of parallel cores}~\cite{yavuz2016genn}
\revision{}{We remark that} the large variability in the performance predictions in \figurename~\ref{fig_parameters}A1,B1,C1 can be at least partially explained by our choice of sampling strategy.
We \revision{decided to }{}generate random numbers of neurons and ranks following a log-uniform distribution, with the effect that all orders of magnitude are equally likely, thus introducing a very large variability.

\paragraph{Minimum network delay affects the relative importance of hardware features}
Another parameter of interest is the minimum network delay, denoted $\delta_{min}$.
In terms of communication, $\delta_{min}$ affects the number of times that global communication must happen to simulate one second of biological time, although it doesn't affect the total number of spikes communicated.
In terms of computation, assuming that the loop ordering strategy to minimize pressure on the memory bandwidth is employed then $\delta_{min}$ affects the number of time iterations in which data locality can be exploited.
\figurename~\ref{fig_parameters}B1 shows the predictions of our performance model for the three in silico models, for different values of the minimum network delay.
Since this delay can only be an integer multiple of the timestep, we exploit the concept of coupling ratio to define a range of plausible minimum delay values by setting a range of values for the coupling ratio and obtaining the corresponding $\delta_{min}$ by multiplication with $\Delta t$.
By looking at the breakdown of performance, we see that larger $\delta_{min}$ improves the performance of inter-node communication but also, somewhat surprisingly, on-node computation.
However, while the communication performance seems to improve indefinitely, the improvement of on-node computation saturates quite quickly.
For the point neuron models, within the range of $\delta_{min}$ values considered here, there is a transition from a regime dominated by communication to one dominated by computation, while the Reconstructed model is dominated by computation for all $\delta_{min}$ values.
To investigate the reasons for changes in performance, we plot the breakdown of relative importance of different hardware features in \figurename~\ref{fig_parameters}B1,B2,B3.
In the case of COBA models, larger $\delta_{min}$ correspond to decreased pressure on the memory bandwidth and network latency, and larger pressure on more scalable hardware features such as CPU instruction throughput and caches throughput.
This points to the fact that simulations based on COBA models with a large $\delta_{min}$ could strongly benefit from shared-memory parallelism.
On the other hand, a larger $\delta_{min}$ in the CUBA model results in decreased pressure on the network latency, but a higher pressure on memory bandwidth.

\paragraph{Large fan-in can be advantageous for performance of point neuron models, but has almost \revision{}{no} effect on Reconstructed model}
Finally, we examine the effect of fan-in, defined as the average number of incoming connections per neuron and denoted by $K$.
This parameter has subtle effects on performance that are difficult to analyze.
For COBA models, a larger fan-in technically means more synapses to simulate thus an expected degradation of performance.
Additionally, for all models a larger $K$ determines an increase in event-driven computation, thus once again an expected degradation of performance.
These hypothesis are confirmed in \figurename~\ref{fig_parameters}C1, which shows that $K$ does not affect communication but has a very strong effect on the Reconstructed model.
Unexpectedly, fan-in seems to only marginally affect the performance of the Simplified model, in spite of it being a COBA model too.
This can be easily explained by the fixed number of synaptic instances in this model (28 excitatory and 8 inhibitory~\cite{rossert2016automated}), such that much like the CUBA Brunel model, ultimately the fan-in affects only the average number of events a neuron must integrate within a certain time period.
Another important point should be made about the effect of fan-in, because changing the number of connections of a neuron has an impact on the in silico model's memory requirements.
\figurename~\ref{fig_parameters}C2 shows the ratio of neurons that can fit in a Gigabyte of memory, according to our performance model, as a function of fan-in.
In a strong scaling scenario, this information sheds new light on the conclusions above, because if only $\frac{1}{x}$ neurons fit in a GB, this can result potentially in a $x$-fold increase in performance from parallelism (disregarding potential communication bottlenecks).
Therefore for the Brunel and Simplified model it can be advantageous to have a large number of incoming connections per neuron, because the performance price paid is more than compensated by the required increase in parallelism.
Conversely, in the Reconstructed model, these two effects appear to balance out almost evenly.
For very large scale simulations, another subtle effect of fan-in is represented by the size of the connection table, an issue that was raised and investigated in~\cite{kunkel2014spiking}.
The connections table of a given rank contains all the GIDs of presynaptic neurons that are relevant for at least one neuron in that rank.
The size of the connection table depends, among other things, on $K$ as well as the total number of neurons and number of distributed ranks.
In \figurename~\ref{fig_parameters}C3 we plot the values of the expected total size of the connection table on the (total number of neurons, number of distributed ranks)-plane as a contour plot, highlighting the contours corresponding to a total size of 1\KB, 1\MB~and 1\GB.
Although in some in silico models connectivity may be determined by complex rules influenced by cell type and spatial locality, for simplicity we compute here the expected size of the connections table assuming uniform connection probability and we also assume that neurons are randomly distributed across ranks.
In a strong scaling scenario the size of the connections table steadily decreases as the number of ranks increases, starting from the minimum of two ranks required by a distributed simulation.
This can be explained by the fact that fixing the network size and increasing the number of ranks entails that there will be less incoming connections to a given rank.
In a weak scaling scenario, such as the maximum filling regime, the size of the connections table transiently increases when the number of distributed ranks is low, but at large scale reaches a constant value steady state determined only by $K$.

\section*{Conclusions}
In this work we \revision{have analyzed}{have delivered a quantitative characterization of} the performance properties of different published in silico models \revision{on the basis}{at the core} of state-of-the-art brain tissue simulations.
Using a grey-box model that combines biological and algorithmic properties with hardware specifications we have identified performance bottlenecks under different simulation regimes, corresponding to a variety of prototypical scientific questions that can be answered by simulations of biological neural networks.

\paragraph{General purpose computing has sustained a diverse performance landscape up to now}
Our results show that there exists a large diversity of performance profiles and bottlenecks that shape the landscape of brain tissue simulations, corresponding to the diversity of sizes and scales at which research questions in simulation neuroscience can be asked.
Thus, our research highlights that the computational neuroscience community is currently greatly benefitting from the adaptability of general purpose computing, exploiting the ease of development and high performance capability to explore different areas of the modeling landscape.

\paragraph{Memory bandwidth and network latency severely limit maximum filling and real time strong scaling}
Using a state-of-the-art HPC server CPU and cluster as a reference, our analysis revealed that all the in silico models saturate the memory bandwidth using quite a small number of shared memory threads.
Even when algorithmic improvements are put into place to mitigate this effect we have identified that the coupling ratio, a dimensionless number that counts the number of timesteps in a mininum network delay period, strongly regulates the saturation of memory bandwidth and, in the extreme case of the Simplified model analysed here, effectively prevents any benefit to be gained from the effort of developing a more efficient algorithm.
Additionally, we discovered that it is not the level of morphological detail, but rather the formalism used to represent synapses, that is the most important factor in explaining the memory bandwidth saturation profile, with COBA models saturating much faster than the CUBA model.
In distributed simulations we identified the network latency, and not the network bandwidth, as the major bottleneck for scaling to very large networks or very large cluster sizes.
\revision{}{This provides a new motivation and justification for the extensive efforts described in~\cite{navaridas2012analytical} in designing a specific communication infrastructure for the SpiNNaker neuromorphic system.}

\paragraph{Model-specific features have a significant impact on shared-memory performance}
Inspection of our performance model allowed us to pinpoint which kernels, hardware specifications and model parameters have the largest impact on performance.
The Brunel model based on the CUBA formalism and IAF neurons is mainly bounded by the spike delivery kernel, which exhibits a good shared-memory scaling behaviour and, in the case of extreme strong scaling, a strong dependence on the inter-cache data paths for good performance.
The two COBA models we analyzed, i.e. Simplified and Reconstructed, have a similar shared-memory scaling behaviour, mainly driven by the \emph{current} kernels required to compute the contributions of individual synapses (and ion channels) to the membrane potential equation.
However, while the Simplified model is 100\% dominated by memory bandwidth, the morphologically detailed Reconstructed model is dominated partially (around 40\%) by other hardware components such as caches and CPU throughput.
It becomes clear that a performance model and a detailed performance analysis are fundamental tools to disentagle the complex web of relationships between in silico models, their software implementation and hardware concretization.

\paragraph{Static and dynamic model parameters affect performance in significant but subtle ways}
Finally, we examined the impact of model parameters on the performance profiles described above.
We found that firing frequency, but surprisingly also minimum network delay, can have a large impact on determining which hardware features may constitute a performance bottleneck.
For firing frequency it is obvious that larger values correspond to more operations required by the simulation algorithm, and thus a lower performance, but our analysis shows that different values of the firing frequency also change the relative importance of hardware features.
Interestingly we found that the minimum network delay, in spite of it not affecting the total number of operations per simulated second, can have an effect on performance simply by shifting the importance of the hardware bottlenecks.
We also found that the average number of incoming connections per neuron plays a subtle role in influencing performance.
Trivially, a larger fan-in increases the computational requirements of a single neuron.
However, it also increases the memory capacity requirements, thus requring a larger degree of parallelism to handle the same network size.
This creates a tradeoff between performance degradation arising from larger computational requirements and performance improvement from parallelism requirements.

\paragraph{Discussion and future improvements}
In this work we have concentrated solely on the aspect of maximizing performance, without considering limitations such as cost or energy.
However, it must be stated that energy efficiency is a central issue in the computational neuroscience community, and one of the main selling points of neuromorphic hardware~\cite{cassidy2014real,stromatias2013power}.
Therefore, a meaningful extension to this work would be to incorporate a model for power consumption alongside performance prediction, as a way to constrain the feasibility and efficiency of certain simulation configurations.
To achieve this, one could exploit already established power consumption models that are easily integrated with the ECM and have been shown to provide valuable insight into the power and performance properties of simulation kernels~\cite{hager2016exploring,hofmann2018accuracy}.
From the modeling point of view, an important aspect that we have neglected in this analysis is synaptic plasticity.
A large portion of research questions that require brain tissue simulations involve learning and synaptic plasticity, so this represents an important extension to our analysis.
However, in this work we decided to concentrate on the inference part of brain tissue simulations because the diversity and complexity of plasticity models warrants a separate analysis.
Given that the performance modeling infrastructure is already in place, we believe that the addition of plasticity would not be a technical challenge, but it would considerably complexify the resulting analysis.
\revision{Finally, to better understand future trends}{Even though we already considered potential hardware improvements in our analysis},  it would be interesting to extend \revision{our performance model}{this study} to include hardware with different a design \revision{features}{space} such as \revision{}{the non-overlapping caches of} AMD CPUs~\cite{hager2017benchmarking} or \revision{}{massive SIMD parallelism of} GPGPUs~\cite{knight2018gpus}.
\revision{as well as}{Finally, the methods developed in this paper can be extended to} different simulation approaches \revision{}{ranging from different communication strategies~\cite{kozloski2011ultrascalable,ananthanarayanan2007anatomy} up to fully asynchronous executions~\cite{magalhaes2019fully}}.

\paragraph{Performance modeling is required to enable next-generation high performance brain tissue simulations}
Our analysis shows that, \revision{given the current state-of-the-art of general-purpose hardware architectures}{if future iterations of general-purpose hardware architectures maintain the same balance as the current state-of-the-art}, it will be very difficult\revision{, if not downright impossible,}{} to achieve fast, large scale simulations of brain tissue.
Even if hardware peak performance were to improve significantly over the next years, the required speedup could only be achieved via specifically targeted advancements and under very restrictive simulation and model configurations.
To support the next generation of brain tissue simulations, the community must therefore focus on the design of dedicated hardware.

In this work we show that the task of designing the hardware and software necessary to achieve high performance simulations is made difficult by the fact that neurons are not just neurons, but these exists instead a large diversity within the performance landscape.
The level of diversity means that design tradeoffs will inevitably require very restrictive decisions about the scale, model and configuration of the target simulation, such as e.g.~\cite{navaridas2012analytical,knight2018gpus}.
For example, while the high-bandwidth delivered by GPGPUs can provide significant speed-up in simulations of medium to large networks of neurons, our analysis points out that this improvement can be undermined by \revision{}{their extremely large number of parallel cores, which makes them susceptible to} dynamic load imbalance caused by irregular firing patterns (see \figurename~\ref{fig_parameters}), or by static load imbalance caused by different neuronal morphologies in the case of detailed neurons (see \figurename~\ref{fig_metrics}).
These phenomena have been observed empirically in the literature, by Yavuz \emph{et al.}~\cite{yavuz2016genn} and Kumbhar \emph{et al.}~\cite{kumbhar2016coreneuron} respectively.
To give another example, in the context of accelerating small networks of neurons up to real-time, we predict that CUBA models would benefit from architectures with a simple but fast cache hierarchy, while for COBA models the cache bandwidth \revision{and}{as well as} the throughput of CPU arithmetic operations must be improved in parallel in order observe the desired speedup.
Additionally, simulation parameters can influence the performance profile in a tangible way.
In the case of the Simplified model, for example, simulations are on average bounded by memory bandwidth and network latency at the current value of the minimum network delay, but the profile would change drastically if the $\delta_{min}$ were increased to the same value of the Brunel model, with memory bandwidth losing importance in favour of cache and CPU throughput (see \figurename~\ref{fig_parameters}).
Even in ideal cases where the in silico experiment falls perfectly within the design space of the hardware and software being used, simulation dynamics outside of the control of the modeler such as firing frequency can rapidly push the performance of simulations towards suboptimal regimes.

Ultimately, we stress that the diversity and complexity of the performance landscape warrants the need for performance modeling to achieve high performance simulations of brain regions and eventually the whole brain.
We believe that our work embodies a concrete step in defining and understanding key performance properties of a wide variety of in silico models, necessary to enable the next generation of brain tissue simulations.


\section*{Methods}
\subsection*{Single-node performance model}
\paragraph{The Execution-Cache-Memory model}
The Execution-Cache-Memory (ECM) model is used to obtain runtime predictions at the granularity of individual clock cycles.
The applicability of the ECM model to brain tissue simulations has been demonstrated in the context of morphologically detailed neurons in~\cite{cremonesi2019analytic}, and this work represents an extension of that analysis.

The ECM model, introduced by~\cite{treibig2010introducing} and refined in~\cite{sthw15,hofmann2018accuracy}, is based on computing two potential bottlenecks separately, then making a prediction based on the slowest one.
The two bottlenecks are the ``overlapping'' execution time, represented by the execution of code in the core, and the data-traffic time.
Several assumptions are required in order to reduce the complexity of estimating these two contributions.
For code execution in the core, we assume that either all instructions in a loop are executed at their maximum throughput, or a Critical Path (CP) dominates the runtime.
In both cases we make extensive use of the IACA~\cite{IACA} analysis tool provided by Intel to obtain reasonably accurate predictions.
For estimating data-traffic time, we combine the knowledge about the speed of transfers between caches provided by vendors with the empirical observation that, on all recent Intel server microarchitectures, the best accuracy in predictions is obtained assuming that there is no temporal overlap between any cache transfers~\cite{treibig2010introducing}.
Finally, Intel's Skylake architecture requires some extra care in modeling cache transfers because of the L3 victim cache setup, for which an ECM model was never previously published.
In this work, we found that the assumptions that both clean and dirty cache lines evicted from L2 are copied in L3, while all read traffic from DRAM goes straight to L2, lead to the most accurate predictions.
An in-depth analysis yielding satisfactory accuracy in the predictions has been presented in~\cite{hager2018applying,cremonesi2019analytic}.

\paragraph{Reference architecture: Intel Skylake-X}
Even though the methodology proposed in this paper is general, to obtain concrete performance predictions and bottleneck analysis we are required to focus on a target architecture.
To uphold a satisfactory level of relevance for the high performance computing community as well as generalizability to future architectures, we picked a modern, general-purpose Intel server architecture, a Intel Skylake (SKX) Intel(R) Xeon(R) Gold 6140 with AVX512 vectorization and Sub-NUMA clustering turned off.
The most relevant hardware characteristics required by the performance model are summarized in Table~\ref{tab_hw}, but we refer to~\cite{cremonesi2019analytic} for the full details.
We obtained these values either directly from the vendor's spec sheets, by custom-designed benchmarks or from the reference tables in~\cite{fog2017instruction}.

\begin{table}[ht!]
\begin{adjustwidth}{-0.25in}{0in} 
\centering
\caption{
{\bf Hardware characteristics of reference architecture SKX-AVX512.}}
\label{tab_hw}
\begin{tabular}{|l|c|c|}
    \hline
      & value & unit \\ \thickhline
      CPU freq & 2.3 & GHz \\ \hline
      Peak DP performance & 1324.8 & \GFS \\ \hline
      Mem BW  & 105 & GB/s \\ \hline
      L1-L2 BW per core & 64 & \BC \\ \hline
      L2-L3 BW per core & 2 $\times$ 16 & \BC \\ \hline
      vector \texttt{exp()} throughput & 1.5 & \CIT \\ \hline
      scalar \texttt{exp()} latency & 22.2 & \CIT \\ \hline
\end{tabular}
\end{adjustwidth}
\end{table}

\paragraph{Simulation of in silico models and experiments}
We used the CoreNEURON software presented in~\cite{kumbhar2019coreneuron} to benchmark and validate the in silico models and experiments analyzed in this paper.
CoreNEURON is a biological neural network simulation tool based on the compute engine of the NEURON simulator, optimised both for memory usage and computational speed.
Both NEURON and CoreNEURON allow scientists to expand the repertoire of available ion channel, synapse, point neuron models and stimuli by using a domain-specific language called NMODL.
A source-to-source compiler is then in charge of translating files from NMODL to C, to be loaded as a dynamic library by the simulation engine at execution time.

For the Reconstructed model, a NEURON implementation was available for download from the Blue Brain NMC portal introduced in~\cite{ramaswamy2015neocortical}.
We chose a cortical layer 5 thick-tufted pyramidal cell as a representative example (specifically the L5\_TTPC1\_cADpyr232\_1 model), because it is one of the largest cells in the reconstructed circuit and contains several different types of ion channels.
In order to maintain generality, we reversed a hand-tuned optimization that comes in the downloaded code for the synaptic state update: it exploits the fact that the differential equations describing the state update are extremely simple to bypass the code generation step, forcing it to use a precomputed decay value while avoiding the call to the expensive exponential function.
However, we believe this optimization would significantly hinder the generality of our results, thus we rolled back the hand-tuned code and reverted to a standard definition of the synaptic state differential equations, leading to the presence of several calls to the exponential function in the kernel.
For the ion channel state and ion channel current kernels, to avoid complexity and gain generality we decided to compute an average of the different ion channel types in the neuron model, weighted by the number of compartments where that ion channel type is present.
The linear algebra kernel requires an in-depth analysis because its structure prevents vectorization, see~\cite{cremonesi2019analytic}.
A few additional tweaks to improve readability and allow better benchmarking were made, but none of them impacted the instruction count nor the computational properties of the kernels.
We refere the interested reader to~\cite{cremonesi2019analytic} for more details on the ECM model for individual ion channel and synapse models.

For the Simplified model, a NEURON implementation was provided to us by the authors of~\cite{rossert2016automated}.
The Simplified model is based on the concept of simplification of the Reconstructed neural microcircuit, by reducing the morphologically detailed neurons to COBA GIF point neuron model as described in~\cite{pozzorini2015automated}.
In the Simplified model all synaptic connections between neurons are ``routed'' through one of 36 synaptic processes on each neuron (28 excitatory and 8 inhibitory), depending on their transmission delay and physiological properties.
This allows a significant reduction of memory and computational footprints.
Since the synaptic model is based on that of the Reconstructed model, we performed the same optimization rollback described above to maintain generality in our discussion and results.

For the Brunel model, we implemented the kernels ourselves using the NMODL language, based on the reference implementation iaf\_psc\_alpha.cpp in the NEST software.
The underlying neuron model is the integrate-and-fire model with alpha-shaped current~\cite{gerstner2014neuronal}.
Since the Brunel model is based on the CUBA formalism, it does not contain any current kernels, although it still has a PSC kernel in which the total current contributions from synaptic events is added to the equations governing the membrane potential dynamics.

\paragraph{ECM model of clock-driven kernels and validation}

We computed the ECM model for all the relevant kernels constituting the simulation workflow presented in Fig~\ref{fig_metrics}.
We present the results in Table~\ref{tab_ecm}, following the notation introduced in~\cite{treibig2010introducing}:
\begin{equation*}
\begin{split}
    T_{contributions}~~=&~~\ecm{T_{OL}}{T_{nOL}}{T_{L1L2}}{T_{L2L3}}{T_{L3Mem}}{\cycles} \\
    T_{predictions}  ~~=&~~\ecmp{T^{ECM}_{L1}}{T^{ECM}_{L2}}{T^{ECM}_{L3}}{T^{ECM}_{Mem}}{cy}.
\end{split}
\end{equation*}
The units for all runtime predictions are cycles per scalar iteration, denoted as [cy], while the units for performance are Giga-scalar iterations per second, denoted by [Giga/s].

\begin{table}[ht!]
\begin{adjustwidth}{-0.25in}{0in} 
\centering
\caption{
{\bf ECM model of clock-driven kernels.}}
\label{tab_ecm}
\begin{tabular}{|l|c|c|}
    \hline
    kernel & model [\cycles] & predictions [\cycles] \\ \thickhline
syn current & \ecm{ 7.20 }{ 3.50 }{ 3.21 }{ 8.33 }{ 4.50 }{} & \ecmp{ 7.20 }{ 7.20 }{ 15.04 }{ 19.54 }{} \\ \hline
ion channel current & \ecm{ 4.68 }{ 2.56 }{ 1.92 }{ 5.07 }{ 2.70 }{} & \ecmp{ 4.68 }{ 4.68 }{ 9.55 }{ 12.25 }{} \\ \hline
linear algebra & \ecm{ 8.10 }{ 6.00 }{ 1.40 }{ 4.00 }{ 1.90 }{} & \ecmp{ 8.10 }{ 8.10 }{ 11.40 }{ 13.30 }{} \\ \hline
syn state & \ecm{ 9.70 }{ 1.70 }{ 1.50 }{ 4.00 }{ 2.10 }{} & \ecmp{ 9.70 }{ 9.70 }{ 9.70 }{ 9.70 }{} \\ \hline
ion channel state & \ecm{ 15.82 }{ 2.16 }{ 1.59 }{ 3.56 }{ 2.24 }{} & \ecmp{ 15.82 }{ 15.82 }{ 15.82 }{ 15.82 }{} \\ \hline
filtered synapse current & \ecm{ 7.44 }{ 3.50 }{ 3.51 }{ 9.03 }{ 4.92 }{} & \ecmp{ 7.44 }{ 7.44 }{ 16.03 }{ 20.95 }{} \\ \hline
gif current & \ecm{ 9.88 }{ 3.38 }{ 3.75 }{ 11.00 }{ 5.26 }{} & \ecmp{ 9.88 }{ 9.88 }{ 18.12 }{ 23.38 }{} \\ \hline
filtered synapse state & \ecm{ 13.31 }{ 2.62 }{ 2.00 }{ 5.50 }{ 2.80 }{} & \ecmp{ 13.31 }{ 13.31 }{ 13.31 }{ 13.31 }{} \\ \hline
gif state & \ecm{ 28.50 }{ 6.25 }{ 2.38 }{ 6.50 }{ 3.33 }{} & \ecmp{ 28.50 }{ 28.50 }{ 28.50 }{ 28.50 }{} \\ \hline
iaf update & \ecm{ 2.41 }{ 1.75 }{ 2.62 }{ 8.00 }{ 3.68 }{} & \ecmp{ 2.41 }{ 4.38 }{ 12.38 }{ 16.05 }{} \\ \hline
iaf psc & \ecm{ 0.62 }{ 0.38 }{ 1.25 }{ 3.00 }{ 1.75 }{} & \ecmp{ 0.62 }{ 1.62 }{ 4.62 }{ 6.38 }{} \\ \hline
\end{tabular}
\end{adjustwidth}
\end{table}

\begin{figure}[ht!]
    \includegraphics[width=0.95\textwidth]{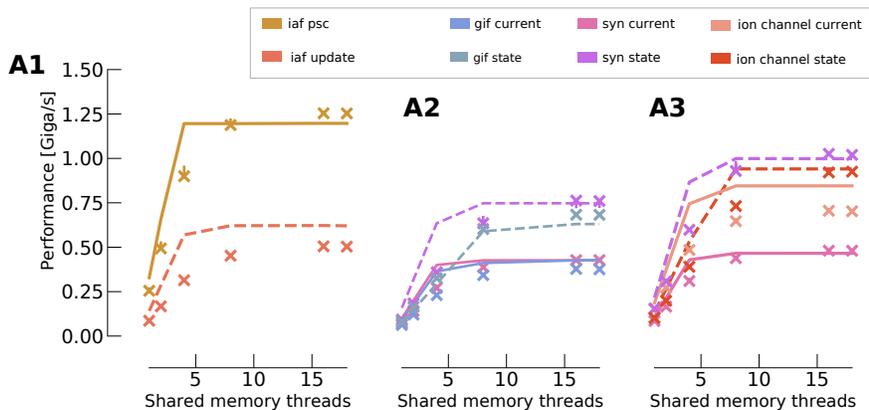}
    \caption{{\bf Validation of performance modeling of in silico models and experiments.}
    Lines represent our predictions using the ECM model, while X markers represent median benchmark measurements and error bars represent the 25\%-75\% percentiles.
    To improve readibility we used dashed lines for state update kernels and solid lines for current kernels.
    All benchmarks were designed with big enough datasets to ensure data was always coming from DRAM.
    {\bf A1,A2,A3} Computational kernels of the Brunel, Simplified and Reconstructed model respectively.
}
\label{fig_valid}
\end{figure}

To validate our predictions we benchmarked and measured the performance of the individual kernels in a simulation that is representative of a typical workload.
For simplicity, in all the benchmarks presented here we made sure that the dataset was large enough to only fit in DRAM, however a more general validation was conducted in~\cite{cremonesi2019analytic} for the kernels constituting the Reconstructed model.
For all in silico models we validated our predictions for the serial execution as well as for shared memory scaling.
Validation results for the Brunel, Simplified and Reconstructed model are presented in Table~\ref{tab_valid} and Fig~\ref{fig_valid}A1,A2,A3 respectively.
In the serial case, the prediction errors are all within 20-30\% of the measured runtime.
Obtaining more accurate predictions is not an easy task, and the reasons for this can vary across kernels.
For ion channel state and current kernels, our model contains an inevitable bias since it represents an average of different ion channel types.
For the linear algebra kernel, a large variability was observed in the serial case.
While it is still possible to obtain reasonably accurate predictions for the state kernels, it must be noted that ultimately it is extremely hard to predict the dynamic behaviour of the out-of-order engine in a complex, modern architecture.
Finally, given that most of the predictions are optimistic, it is reasonable to assume that performance limiting factors such as dynamic CPU throttling, as well as intrinsic factors such as critical paths and instruction latencies, could be impacting the performance negatively.
Digging deeper into the complexity of the computer architecture to obtain more accurate predictions is outside the scope of this paper, whose purpose is to employ performance modeling as a means of explaining bottlenecks and predicting future trends, not to implement optimizations at the granularity of individual cycles.

\begin{table}[ht!]
\centering
\caption{
{\bf Validation of ECM performance model for all kernels.}
Measurements are shown as median values $\pm$ inter-quantile range from a dataset of 10 independent benchmark executions.}
\begin{tabular}{|l|r|r|r|r|}
\hline
 & \multicolumn{2}{c|}{serial} & \multicolumn{2}{c|}{parallel} \\ \hline
 kernel              &   pred [\cycles] & meas [\cycles]   &   pred [\cycles] & meas [\cycles]   \\ \thickhline
\hline
 syn current                  &   19.5   & 24.6$\pm$1.5   &  4.5   &   4.4$\pm$0.1   \\ \hline
 ion channel current          &   12.2   & 15.2$\pm$0.3   &  2.7   &   3.3$\pm$0.1   \\ \hline
 linear algebra               &   13.3   & 18.8$\pm$5.3   &  1.9   &   2.2$\pm$0.2   \\ \hline
 syn state                    &    9.7   & 13.8$\pm$0.2   &  2.1   &   2.0$\pm$0.0   \\ \hline
 ion channel state            &   15.8   & 20.6$\pm$0.1   &  2.2   &   2.3$\pm$0.0   \\ \hline
 filtered synapse current     &   21.0   & 28.9$\pm$1.8   &  4.9   &   4.9$\pm$0.1   \\ \hline
 gif current                  &   23.4   & 33.7$\pm$1.7   &  5.3   &   6.0$\pm$0.2   \\ \hline
 filtered synapse state       &   13.3   & 21.1$\pm$0.4   &  2.8   &   2.8$\pm$0.1   \\ \hline
 gif state                    &   28.5   & 25.4$\pm$0.0   &  3.3   &   3.1$\pm$0.0   \\ \hline
 iaf update                   &   16.1   & 26.4$\pm$0.9   &  3.7   &   4.5$\pm$0.0   \\ \hline
 iaf psc                      &    6.4   & 8.2$\pm$0.6    &  1.8   &   1.7$\pm$0.0   \\ \hline
\end{tabular}
\label{tab_valid}
\end{table}
\paragraph{Spike delivery kernel}
The spike delivery kernel is arguably the most distinguishing aspect of brain tissue simulations, when compared to models of other biological or physical phenomena.
Regarding its performance aspects, we have shown that this kernel has a very different impact on overall performance for CUBA and COBA models.

In terms of algorithm design, all state-of-the-art software use some sort of priority queue or ring buffer to store synaptic events to be delivered within a timestep.
For benchmarking, we have decided to separate the operations related bookeeping of events inside the queue from the actual kernel execution, and we only consider the latter because the scope of this paper is restricted to computational and communication kernels.
We slightly changed the benchmarking approach from~\cite{cremonesi2019analytic} because we found that, given the implementation infrastructure in CoreNEURON, the runtime overhead
due to programming design was comparable to the runtime of a CUBA kernel.
Therefore we first tried to estimate this overhead by executing an \emph{empty} kernel and subtracted the resulting overhead from the actual benchmarks.

The spike delivery kernel is characterized by erratic memory accesses, because the order of activation of synapses is unpredictable.
In this paper we always consider the worst possible case in which every spike to be delivered could not be cached and thus must come from main memory.
In~\cite{cremonesi2019analytic} we have shown that the synapses in the Reconstructed model, and therefore also the Simplified one since they have the same delivery kernel, were affected by the latency of the memory system but not completely dominated by it.
The reason is that while the CPU is handling the delivery of one spike, it can ``look ahead'' and schedule requests for the data of the next one.
This is different from the classic purely latency bound kernels in which the CPU is allowed to begin an iteration only after the previous is fully completed.
On the other hand, all the requests for data are noncontiguous and therefore none of the pipelining and prefetching techniques are very effective in hiding the latency of fetching the data.

\begin{figure}[ht!]
    \includegraphics[width=0.95\textwidth]{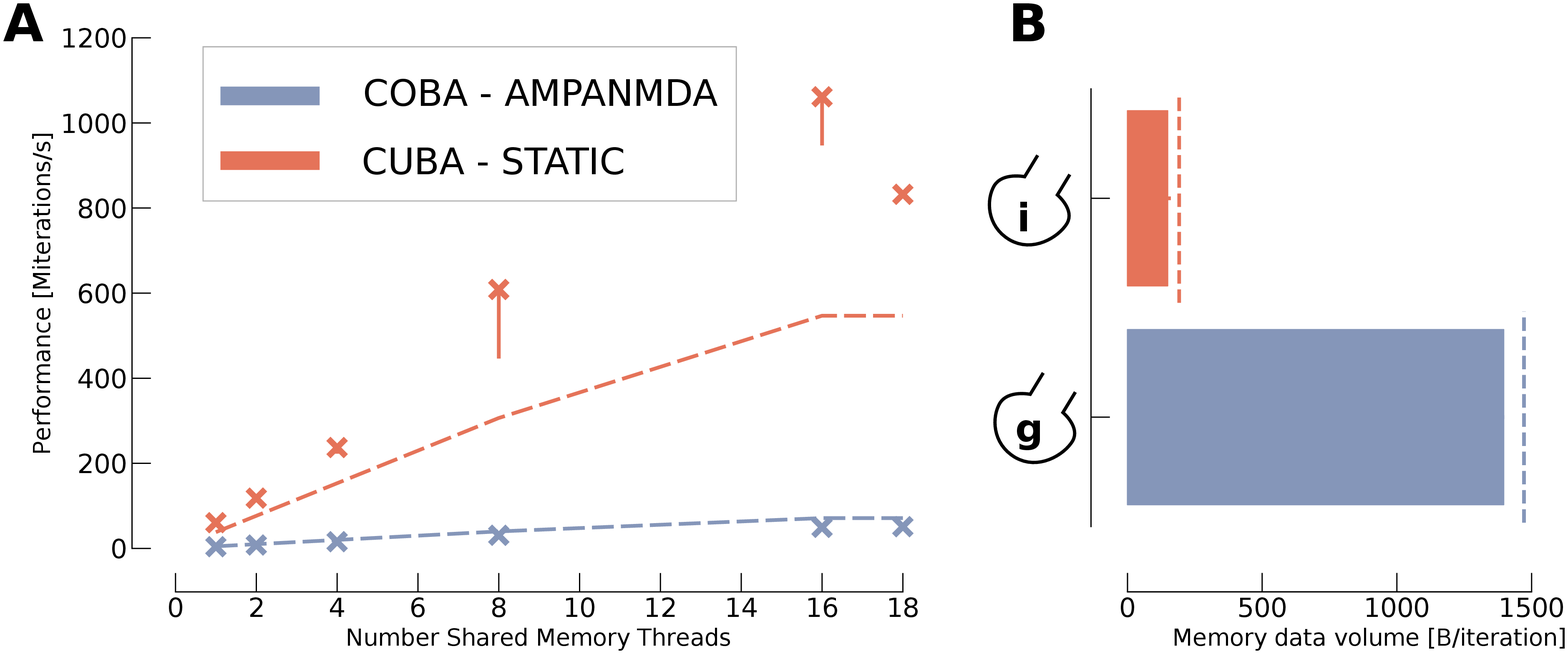}
    \caption{{\bf Validation of the spike delivery kernel.}\\
{\bf A} Performance predictions (dashed lines) and benchmark measurements (X markers).
Performance is measured in $10^6$ spikes delivered per wallclock second.
Error bars represent the 25\% and 75\% percentiles.
{\bf B} Measured memory traffic per delivered spike in Bytes.
Dashed lines represent the predictions from our model.}
\label{fig_valid_delivery}
\end{figure}

For the COBA delivery kernel, we found that the strategy of designing a synthetic benchmark with a similar access pattern to estimate an adjusted latency worked very well.
On our validation architecture, the adjusted latency was around 20 \cycles per memory access.
This led us to a single-thread prediction of 460 \cycles per delivered spike, to be compared to the benchmark measurement of 571.6 $\pm$ 14.9 \cycles, which represents roughly a 20\% error.
For multiple threads, we assumed the performance scales linearly with the number of threads until the bottleneck of memory bandwdith is exhausted.
It should be noted in this case that the predicted memory traffic is quite large because we assume that for every memory access, a full cache line (64 B) must pulled from the memory even though only a single double-precision variable (i.e. 8 B) is required.
At maximum number of threads, this amounts to a predicted runtime of 32 \cycles per delivered spike, against  benchmark measurements of 45.0 $\pm$ 0.1 \cycles, i.e. a 28\% error.
Applying the same strategy to the CUBA delivery kernel yielded acceptable but not perfect predictions.
For the single thread, we predict a runtime of 60 \cycles but measured 38.0 $\pm$ 1.8 \cycles, giving an error of 58\%, while at maximum thread we predicted a runtime of 4 \cycles but measured 2.8 $\pm$ 0.04, i.e. a 43\% error.
Given our current understanding of this kernel, and the complexity introduced by the out-of-order execution and memory access scheduling, we deem these predictions quite satisfactory.
In \figurename~\ref{fig_valid_delivery} we present the predicted and measured performance and memory traffic for the spike delivery kernel.
\paragraph{ECM model predictions and inference}
During our analysis, we extensively used the ECM model to perform inference on the performance of in silico models and experiments.
For completeness we report here our strategy for computing derived metrics or quantities from the raw output of the ECM model.
To compute the memory utilization, shown e.g. in Fig~\ref{fig_nsatur}A, we first computed the memory bandwidth as the ratio between the total memory traffic in a timestep divided by the predicted runtime to execute that timestep.
Then we divided the memory bandwidth by the theoretical peak bandwidth of the system to obtain the memory utilization.\\
Another derived quantity that we make extensive use of is the concept of hardware bottleneck.
To define this, we first introduce the concept of hardware contribution as the sum of all the predicted runtimes in which a particular hardware feature was the determining factor for performance, as shown in Fig~\ref{fig_hw_contrib}.
To give an example, let us take the ECM model for the synaptic current, expressed in notation as: \ecm{ 7.20 }{ 3.50 }{ 3.21 }{ 8.33 }{ 4.50 }{\cycles}.
Taking the case where the data fits in L2, the ECM prediction corresponds to $T^{ECM}_{L2} = T_{OL} = 7.20 $ \cycles, which are all spent in the core executing arithmetic operations.
So even though $3.21$ \cycles~are spent in transfers between the L1 and L2 cache, we do not consider this because the arithmetic operations in the core are the only determining factor.
Eventually, we can refine this metric by splitting $7.20$ \cycles~into $1.5$ \cycles~for the exponential (per scalar iteration, assuming full throughput) and $5.7$ \cycles~for the rest of the arithmetic operations.
Thus the contributions to the synaptic current kernel, when data is in L2, would be exp: $1.5$ \cycles~and CPU: $5.7$ \cycles.
Conversely, when data is in DRAM, the ECM prediction for a serial execution is given by $T^{ECM}_{Mem} = T_{nOL} + T_{L1L2} + T_{L2L3} + T_{Mem} = 19.54$ \cycles.
Therefore in this case, since the only dominating factors correspond to the data transfer, the hardware contributions consist of load: $3.5$ \cycles, L1$\to$L2: $3.21$ \cycles, L2$\to$L3: $8.33$ \cycles, DRAM bw: $4.5$ \cycles.
Considering a parallel execution using 18 threads, if we assume full memory bandwidth saturation then the only dominating factor for performance is the memory bandwidth, and therefore the hardware contributions would be only DRAM bw: $4.5$ \cycles.
Finally, the situation becomes more complicated if we wish to take into account the special loop ordering to avoid bandwidth saturation.
In this case, taking again the example of the synaptic current kernel in the detailed model, in the first out of four time iteration data comes from memory, and in the following three out of four time iterations data comes from L3.
Thus the hardware contributions, averaged over time iterations, would be load: $\frac{3}{4}\times \frac{3.5 \cycles}{18} = 0.15$ \cycles, L1$\to$L2: $\frac{3}{4}\times \frac{3.21 \cycles}{18} = 0.13$ \cycles, L2$\to$L3: $\frac{3}{4}\times \frac{1.57 \cycles}{18} = 0.35$ \cycles, DRAM bw: $\frac{1}{4}\times 4.5 \cycles=1.13$ \cycles.
\subsection*{Interprocess communication performance model}
\paragraph{Spike exchange in brain tissue simulations}
All the in silico models and experiments considered in this paper are based on the Bulk Synchronous Parallel (BSP) model~\cite{valiant1990bridging}, which prescribes a clear distinction between an \emph{on-node computation} phase (happening in a distributed parallel fashion) and an \emph{inter-node communication} phase.
For brain tissue simulations, the inter-node communication phase corresponds to the spike exchange step in Fig~\ref{fig_metrics}B1,B2.
Moreover, we make the assumption that the distributed processing is implemented in MPI, because it represents the current state of practice in the HPC community.
In all the widely-used state-of-the-art simulators, the spike exchange step is implemented by a blocking collective call, typically to a variant of the Allgather operation.
This entails that all the parallel ranks have, at the end of the communication step, knowledge of all the spikes produced by the simulation during the last min-delay period.
Recent work has shown that at extremely large scales, this implementation can become prohibitively expensive in terms of memory requirements, and proposed to use instead the Alltoall operation to deliver spikes only to the ranks where they are required~\cite{jordan2018extremely}.
Other alternative implementations have been suggested, using nonblocking point-to-point communication~\cite{ananthanarayanan2007anatomy} or spatial decomposition~\cite{kozloski2011ultrascalable}.
All these fall outside of the scope of this paper, which is focused on medium-to-large cluster sizes and well established, widely used software solutions.

\paragraph{The LogGP model}
We use the LogGP model~\cite{alexandrov1997loggp} to predict and explain the performance of the spike exchange simulation step.
The LogGP model is an extension to the LogP model~\cite{culler1993logp} that uses an additional parameter, the gap per byte denoted $G$, as a way to account for the sending and receiving of long messages.
The main features of all models based on LogP is that their parameters are easily relatable to hardware characteristics, thus allowing a high degree of interpretability.
In the LogGP model, a distinction is made between two bottlenecks: the processor overhead and the networking hardware.
The $L,g,G$ parameters are all related to the networks and represent respectively an upper bound on the latency, the minimum gap between messages (i.e. the reciprocal of the maximum injection rate) and gap per byte (i.e. the reciprocal of the network bandwidth).
On the processor side, even though the original model prescribes a single parameter $o$ representing a fixed overhead, we followed the approach in~\cite{hoefler2009loggp} and considered two parameters $o_i, o_s$ that represent the intercept and slope of the linear relationship between message size and processor overhead.
Finally, the last parameter of relevance is $P$, i.e. the total number of parallel processes.

\paragraph{Reference architecture: Infiniband EDR with HPE-MPI}
Even though the performance modeling tools considered here can generalize to several types of architectures~\cite{hoefler2009loggp}, to validate our performance predictions we restrict our focus to a representative example of high performance network architecture: a vendor (HPE) provided MPI implementation, based on MPT 2.16 and the MPI 3.0 standard, over an Infiniband EDR 100 GB/s fabric.
We used the Netgauge v2.4.6 tool introduced in~\cite{hoefler2007low} to make a first assessment of the LogGP parameters, and found that according to the recommended interpretation of the output from the tool, two different sets of parameters should be used for small (less than 16kB) and large message sizes.
We report the parameters in Table~\ref{tab_loggp}.

\begin{table}[ht!]
\centering
\caption{
{\bf LogGP parameters.}}
\begin{tabular}{|l|r|r|r|}
\hline
& small sizes & large sizes & unit  \\ \thickhline
$L$          & 1.54  & 1.54   & \us  \\ \hline
$o_i$        & 0.133 & 0.0249 & \us  \\ \hline
$o_s$        & $4.59\times10^{-5}$ & $6.48\times10^{-5}$ & \usB \\ \hline
$g$          & 0.526 & 6.12 & \us \\ \hline
$G$          & $1.42\times10^{-4}$ & $2.07\times10^{-4}$ & \usB \\ \hline
\end{tabular}
\label{tab_loggp}
\end{table}

The LogGP framework allows us to directly model the latency of point-to-point communication using the formula $ (L + 2o_i) + (G + 2o_s)m$, where $m$ is the message size.
However, additional work is required to model collective communication because different algorithms can be used to disseminate the messages across the network.
For the Allgather operation there are several implementations available~\cite{thakur2005optimization}, and the decision of which one to use can be a complex function of the static network characteristics such as the topology as well as dynamic specifications such as the message size and number of ranks involved.
In this work we consider only the ring algorithm, because it is the mosts commonly used~\cite{thakur2005optimization} (especially for large message sizes) and we wish to keep our analsys simple.
Therefore, the total latency to perform an Allgather operation among $P$ parallel ranks with a total message size of $m$ is:
\begin{equation*}
T_{Allgather} = (P-1)(L + 2o_i) + \frac{P-1}{P}(G + 2o_s)m.
\end{equation*}

\paragraph{Validation and inference}

Firstly, it must be noted that the original CoreNEURON implementation of the spike exchange algorithm uses an MPI custom datatype to represent a spike as an aggregate of a double-precision variable representing the time of spike and an integer variable representing the ID of the source neuron.
This implementation, however, can turn out to be unsatisfactory in terms of performance, in particular when one tries to predict the runtime of a collective communication, because hidden data shuffling and copy operations can take place~\cite{carpen2017expected}.
To address the performance issues deriving from custom datatypes, we reimplemented the spike exchange operation to perform two consecutive MPI\_allgatherv operations: the first on the array of timings and the second on the array of source neuron IDs.
To measure the latency of the communication itself, we executed a simulation where we artificially filled every communication buffer on each distributed rank with the same number of spikes, and measured the communication time per rank.

\begin{figure}[ht!]
    \includegraphics[width=0.95\textwidth]{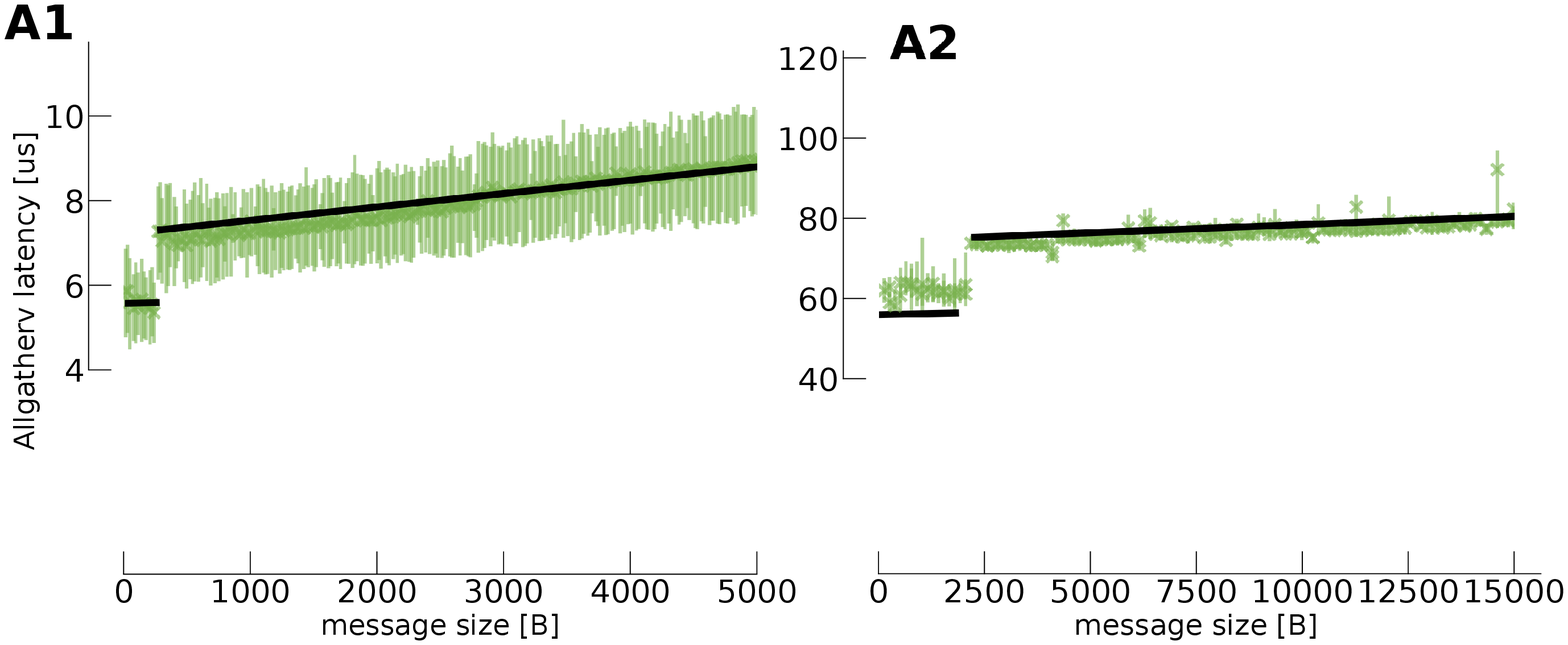}
    \caption{{\bf Validation of interprocess communication performance modeling.}\\
    Latency of the Allgatherv operation in $\mu s$, defined as the median latency across all ranks involved in the communication, for different values of the total message size.
    The error bars represent the maximum and minimum latency over all parallel ranks.
    The black lines represent the prediction from our implementation of the LogGP model.
    To total number of ranks involved in the communication in {\bf A1} was 4, and in {\bf A2} was 32.
}
\label{fig_valid_comm}
\end{figure}

In the process of validation, we discovered that the raw LogGP model based on the small-messages parameters from Netgauge was only valid in the context of very small messages, while the large-messages parameters never seemed relevant for our use case.
In particular, we found that for messages larger than $P \times 65$\bytes~the Allgather operation incurred a penalty in both latency and bandwidth.
The source of this penalty is unclear, as it could be due to a switch in point-to-point protocols, a change in the underlying algorithm or additional communication to ensure synchronization.
More refined performance models exist such as the LogGPS~\cite{ino2001loggps} that try to take global synchronization latencies into account, but we found that a simpler yet effective way to improve our model was to introduce two penalty terms: a latency penalty $p_L = 0.593$\us~and a bandwidth penalty $p_G = 1.875\times10^{-4}$\usB.
In light of this the formula to obtain a prediction for the latency of the Allgather operation becomes:
\begin{equation*}
T_{Allgather} =
\begin{cases}
(P-1)(L + 2o_i) + \frac{P-1}{P}(G + 2o_s)m ~&~ m < 65P\\
(P-1)(L + 2o_i + p_L) + \frac{P-1}{P}(G + 2o_s + p_G)m ~&~ m \geq 65P.
\end{cases}
\end{equation*}
We show in Fig~\ref{fig_valid_comm}B1,B2 the measured and predicted $T_{Allgather}$ for $P=4$ and $P=32$ respectively.
The model has a good overall agreement with the measured data, rarely exceeding 10\% relative error and often within the intrinsic variability of the data itself.
This grants a significant degree of confidence in our predictions, at least for cluster sizes that remain within reasonable distance from our maximum validation size of 32 nodes.

Several shortcomings that can impact the modeling of MPI communication runtimes have been reviewed in~\cite{hoefler2010toward}.
Although our application of the LogGP model addresses some of them, namely the custom datatypes and the overlapping of communication and computation, others such as synchronization, congestion and topology may negatively affect the accuracy of our predictions, especially at very large cluster sizes.
For example, one of the main points advocating for a multicast spike communication strategy presented in~\cite{navaridas2012analytical} is tightly linked with the average number of hops in the underlying network hardware, i.e. a topological parameter which is not considered by our model.
We believe, however, that within the design space of medium to large clusters considered here our predictions are still able to deliver the sufficient accuracy required by a bottleneck analysis.
\section*{Acknowledgments}
This work has been funded by the EPFL Blue Brain Project(funded by the Swiss ETH board).
The authors gratefully acknowledge the compute resources and support provided by the Erlangen Regional Computing Center (RRZE), and in particular Georg Hager for the fruitful discussions regarding the ECM model and the interpretation of performance predictions.
We are also indebted to the BlueBrain HPC team, and in particular Bruno Magalhaes and Pramod Kumbhar, for helpful support and discussion regarding CoreNEURON.

\end{document}